\documentclass{article}
\usepackage{arxiv}
\usepackage[T1]{fontenc}
\usepackage[latin9]{inputenc}
\usepackage{color}
\usepackage{float}
\usepackage{textcomp}
\usepackage{amsmath}
\usepackage{amssymb}
\usepackage{stmaryrd}
\usepackage{graphicx}
\usepackage[unicode=true,
 bookmarks=false,
 breaklinks=false,pdfborder={0 0 1},backref=false,colorlinks=false]
 {hyperref}
\hypersetup{pdftitle={\@shorttitle},
 pdfauthor={\@shortauthor},
 pdfsubject={\@Subject},
 pdfkeywords={Computer Graphics Forum, EUROGRAPHICS},
   colorlinks,linkcolor=blue,citecolor=blue,urlcolor=blue,    bookmarks=false,    pdfpagemode=UseNone}

\makeatletter

\providecommand{\tabularnewline}{\\}
\floatstyle{ruled}
\newfloat{algorithm}{tbp}{loa}
\providecommand{\algorithmname}{Algorithm}
\floatname{algorithm}{\protect\algorithmname}

\usepackage{amsmath,amssymb,amsfonts}
\usepackage{algorithmic}
\usepackage{graphicx}
\usepackage{textcomp}
\usepackage{xcolor}

\usepackage{tikz,tkz-tab,tkz-graph}

\usepackage{amsthm}

\theoremstyle{plain}
\newtheorem{claim}{Claim}{\bf}{\it}

\usepackage{tikz,tkz-tab,tkz-graph}
\usepackage{relsize}

\usetikzlibrary{arrows,automata}
\usetikzlibrary{positioning, fadings,snakes,calc,fadings,shapes.arrows,shadows,backgrounds}
\usetikzlibrary{shapes.geometric}
\tikzset{
    state/.style={
           rectangle,
  fill=#1!5!white,
           draw=#1, very thick,
           minimum height=2em,
           inner sep=2pt,
           text centered,
           },
    highlight/.style={
           rectangle,
  fill=#1!50!white,
           rounded corners,
           draw=#1, very thick,
           minimum height=2em,
           inner sep=2pt,
           text centered,
           },
coeff/.style={
           circle,
           draw=black, very thick,
           minimum height=2em,
           inner sep=2pt,
           text centered,
           },
ptNode/.style={circle, fill=black,thick, inner sep=2pt, minimum size=0.2cm}	,
square/.style={regular polygon,regular polygon sides=4},
ptNodeSq/.style={square, fill=black,thick, inner sep=2pt, minimum size=0.2cm},	
}

\tikzset{
    invisible/.style={opacity=0},
    visible on/.style={alt=#1{}{invisible}},
    alt/.code args={<#1>#2#3}{%
      \alt<#1>{\pgfkeysalso{#2}}{\pgfkeysalso{#3}} 
    },
  }

\tikzset{fontscale/.style = {font=\relsize{#1}}
    }

\usepackage{colortbl}
\definecolor{lightgray}{gray}{0.9}

\definecolor{S_purple}{RGB}{204, 0, 204}
\definecolor{S_brique}{RGB}{204, 51, 0}
\definecolor{S_petrol}{RGB}{0, 102, 153}
\definecolor{S_green}{RGB}{0, 153, 0}

\makeatother

\title{Exchange-Based Diffusion in Hb-Graphs: Highlighting Complex Relationships}

\author{
Xavier Ouvrard \\
University of Geneva, CERN\\
CERN, 1 Esplanade des Particules, CH-1211 Geneva 23\\
\texttt{xavier.ouvrard@cern.ch} \\
\And
Jean-Marie Le Goff \\
CERN\\
CERN, 1 Esplanade des Particules, CH-1211 Geneva 23 \\
\texttt{jean-marie.le.goff@cern.ch} \\
\And
St\'ephane Marchand-Maillet \\
CUI Batelle A, University of Geneva, Route de Drize, 7, CH-1227 Carouge\\
\texttt{stephane.marchand-maillet@unige.ch} \\
}

\begin{document}
\maketitle
\begin{abstract}
Most networks tend to show complex and multiple relationships between entities. Networks are usually modeled by graphs or hypergraphs; nonetheless a given entity can  occur many times in a relationship: this brings the need to deal with  multisets instead of sets or simple edges. Diffusion processes are useful to highlight interesting parts of a network: they usually start with a stroke at one vertex and diffuse throughout the network to reach a uniform distribution. Several iterations of the process are required prior to reaching a stable solution. We propose an alternative solution to highlighting the main components of a network using a diffusion process based on exchanges: it is an iterative two-phase step exchange process. This process allows to evaluate the importance not only of the vertices   but also of the regrouping level. To model the diffusion process, we extend the concept of hypergraphs that are families of sets to families of multisets, that we call hb-graphs.
\end{abstract}

\textcolor{blue}{This article is an extended version of \cite{ouvrard2018hbgraphdiffusion}
(pre-printed in arXiv:1809.00190v1): the text of the extended version
is in blue, the text in black is the one of \cite{ouvrard2018hbgraphdiffusion}.
All the figures except Figure \ref{Fig:Principle_exchange} have been
either modified or added in this extended version to take into account
the new developments. The contributions of this extended version are:
the proofs of conservation and convergence of the extracted sequences
of the diffusion process, as well as the illustration of the speed
of convergence and comparison to classical and modified random walks;
the algorithms of the exchange-based diffusion and the modified random
walk; the application to a use case based on Arxiv publications.}

\section{Introduction}

Many relationships are more than pairwise relations: entities are
often grouped into sets, corresponding to $n$-adic relationships.
Each of these sets can be viewed as a collaboration between entities.
Hypergraphs naturally represent $n$-adic relations. It has been shown
that facets of an information space can be modeled by hypergraphs
\cite{ouvrard2017hypergraphfmwk}: each facet corresponds to a type
of metadata. The different facets are then linked by reference data
attached to hyperedges within that facet. The step forward is to highlight
important information contained in those facets. This is commonly
achieved in hypergraphs using random walks \cite{zhou2007learning,bellaachia2013random}.
Reference \cite{bellaachia2013random} shows that the weighting of
vertices at the level of the hyperedges in a hypergraph provides better
information retrieval. These two approaches - \cite{zhou2007learning,bellaachia2013random}
- mainly focus on vertices; but as hyperedges are linked to references
that can be used as pivots in between the different facets \cite{dork12PivotPaths:StrollingthroughFacetedInformationSpaces,ouvrard2017hypergraphfmwk},
it is also interesting to highlight important hyperedges. For instance,
in a document database, different metadata can be used to label authors,
author keywords, processed keywords, categories, added tags: the pivots
between the different facets of this information space correspond
to the documents themselves. In the specific case of tags, it can
be important to have weights attached to them if the users are able
to attach tags to documents.

Hyperedge-based weighting of vertices is easier to achieve through
multisets: multisets store information on multiplicity of elements.
We use multisets family over a set of vertices, called hyper-bag-graph
- hb-graph for short - as an extension of hypergraphs. Hb-graph multisets
play the role of the hyperedges in hypergraph: they are called hb-edges.

We want to address the following research questions: ``Can we find
a network model and a diffusion process that not only rank vertices
but also rank hb-edges in hb-graphs?''. We develop an iterative exchange
approach in hb-graphs with two-phase steps that allows to extract
information not only at the vertex level but also at the hb-edge level.

We validate our approach by using randomly generated hb-graphs. The
hb-graph visualisation highlights not only vertices but also hb-edges
using the exchange process. We show that the exchange-based diffusion
process provides proper coloring of vertices with high connectivity
and highlights hb-edges with a normalisation approach - allowing small
hb-edges to have a chance to be highlighted. We apply this approach
to process the metadata contained in the results retrieved by querying
Arxiv through its API in order to visualize the results: we will show
how it can be used to allow further query expansion.

This paper contributes to present an exchange-based diffusion process
that enables not only the ranking of vertices but also of hb-edges.
It formalizes exchanges by using hb-graphs that can naturally cope
with elements multiplicity. It contributes also to a novel visualisation
of this kind of network depicted in each facet of the information
space.

In Section II, the mathematical background and the related work is
given. The construction of the formalisation of the exchange process
is presented in Section IV. Results and evaluation are given in Section
V and future work and conclusion are addressed in Section VI.

\section{Mathematical background and Related work}

\subsection{Hypergraphs}

\label{sec:Background}

A hypergraph $\mathcal{H}=\left(V,E\right)$ over a finite set of
vertices $V=\left\{ v_{1}\,;\,v_{2};\,...\,;\,v_{n}\right\} $ is
defined in \cite{berge1973graphs} as a family of hyperedges $E=\left(e_{1},e_{2},...,e_{p}\right)$
where each hyperedge is a non-empty subset of $V$ and such that $\bigcup\limits _{i=1}^{p}e_{i}=V$.
A hypergraph $\mathcal{H}_{w}=\left(V,E,w_{e}\right)$ is said edge-weighted
if there exists an application $w_{e}:E\rightarrow\mathbb{R^{+*}}$.

In a weighted hypergraph the degree $\deg\left(v_{i}\right)$ of a
vertex $v_{i}$ is defined as: 
\[
d_{i}=\deg\left(v_{i}\right)=\sum\limits _{e_{j}\in E:v_{i}\in e_{j}}w_{e}\left(e_{j}\right).
\]
The volume of $S\subseteq V$ is defined as:
\[
\text{vol}(S)=\sum\limits _{v_{i}\in S}\deg\left(v_{i}\right).
\]

The incident matrix of a hypergraph is the matrix $H=\left[h_{ij}\right]_{\substack{1\leqslant i\leqslant n\\
1\leqslant j\leqslant p
}
}$ of $M_{n\times p}\left(\left\{ 0\,;\,1\right\} \right)$, where $h_{ij}=\begin{cases}
1 & \text{if\,}\,\,v_{i}\in e_{j}\\
0 & \text{otherwise}
\end{cases}$.

Random walks are largely used to evaluate the importance of vertices
in hypergraphs. In \cite{zhou2007learning}, a random walk on a hypergraph
is defined by choosing a hyperedge $e_{j}$ with a probability proportional
to $w_{e}\left(e_{j}\right)$; and within that hyperedge a vertex
is randomly chosen using a uniform law. The probability transition
from a vertex $v_{i_{1}}$ to a vertex $v_{i_{2}}$ is: 
\[
p(v_{i_{1}},v_{i_{2}})=\sum\limits _{j=1}^{p}w_{e}\left(e_{j}\right)\dfrac{h_{i_{1}j}}{d_{i_{1}}}\times\dfrac{h_{i_{2}j}}{\delta_{j}},
\]
where $\delta_{j}=\deg\left(e_{j}\right),\,1\leqslant j\leqslant p$
is the degree of a hyperedge defined in \cite{zhou2007learning} as
its cardinality. This random walk has a stationary state which is
shown to be $\pi=\left(\pi_{i}\right)_{1\leqslant i\leqslant n}$
with $\pi_{i}=\dfrac{d_{i}}{\text{vol}V}$ for $1\leqslant i\leqslant n$
\cite{ducournau2014random}. This process differs from the one we
propose: our diffusion process is done by successive steps from a
random initial vertex on vertices and hyperedges.

Reference \cite{bellaachia2013random} defines a random walk for weighted
hypergraphs using weight functions both for hyperedges and vertices:
a vector of weights is built for each vertex making weights of vertices
hyperedge-based; a random walk similar to the one above is then built
that takes into account the vertex weight. The evaluation is performed
on a hypergraph built from a public dataset of computer science conference
proceedings; each document is seen as a hyperedge that contains keywords;
hyperedges are weighted by citation score and vertices of a hyperedge
are weighted with a tf-idf score. Reference \cite{bellaachia2013random}
shows that a random walk on the (double-) weighted hypergraph enables
vertex ranking with higher precision than random walks using unweighted
vertices. This process differs again from our proposal: our process
not only enables simultaneous alternative updates of vertices and
hb-edges values but also provides hb-edge ranking. We also introduce
a new theoretical framework to perform our diffusion process.

Random walks relate to diffusion processes. Reference \cite{lee2011hyper}
uses random walks in hypergraph for image matching. Reference \cite{lu2011high}
builds higher order random walks in hypergraph and constructs a generalised
Laplacian attached to the graphs generated from their random walks.

\textcolor{blue}{Hypergraphs fit to model multi-adicity in structures
where the traditional pairwise relationship of graphs is insufficient:
they are used in many areas such as social networks in particular
in collaboration networks - \cite{newman2001scientific,newman2001scientific-2}
-, co-author networks - \cite{grossman1995portion}, \cite{taramasco2010academic}
-, chemical reactions - \cite{temkin1996chemical} - , genome - \cite{chauve2013hypergraph}
-, VLSI design - \cite{karypis1999multilevel} - and other applications.
Hypergraphs are also used in information retrieval for different purposes
such as query formulation in text retrieval \cite{bendersky2012modeling},
in music recommandation \cite{bu2010music},... Several applications
of hypergraphs exist based on the diffusion process firstly developped
by \cite{zhou2007learning}. \cite{gao20123} uses \cite{zhou2007learning}
for 3D-object retrieval and recognition by building multiple hypergraphs
of objects based on their 2D-views that are analysed using the same
approach. In \cite{zhu2015content}, multiple hypergraphs are constructed
to characterize the complex relations between landmark images and
are gathered into a multimodal hypergraph that allows the integration
of heterogeneous sources providing content-based visual landmark searches.
Hypergraphs are also used in multi-feature indexing to help image
retrieval \cite{xu2016multi}. For each image, a hyperedge gathers
the first $n$ most similar images based on different features. Hyperedges
are weighted by average similarity. A spectral clustering algorithm
is then applied to divide the dataset into $k$ sub-hypergraphs. A
random walk on these sub-hypergraphs allows to retrieve significant
images: they are used to build a new inverted index, useful to query
images. In \cite{wang2018joint}, a joint-hypergraph learning is achieved
for image retrieval, combining efficiently a semantic hypergraph based
on image tags with a visual hypergraph based on image features.}

\subsection{Multisets}

Multisets - also known as bags or msets - have a long use in many
domains. But before developping their use in different domains, we
firstly give main definitions on multisets mainly based on \cite{singh2007overview}.

A \textbf{multiset} is a pair $A_{m}=(A,m)$ where $A$ is a set of
distinct objects and $m$ is an application from $A$ to $\mathbb{W}\subseteq\mathbb{R}$
or $\mathbb{N}$. $A$ is called the \textbf{universe }of the multiset
$A_{m}$, $m$ is called the \textbf{multiplicity function} of the
multiset $A_{m}$. $A_{m}^{\star}=\left\{ x\in A:m(x)\neq0\right\} $
is called the \textbf{support} of $A_{m}$. The elements of the support
of an mset are called its \textbf{generators}.

A multiset where $\mathbb{W}\subseteq\mathbb{N}$ is called a \textbf{natural
multiset}.

The \textbf{m-cardinality} of $A_{m}$ written $\#_{m}A_{m}$ is defined
as:

\[
\#_{m}A_{m}=\sum\limits _{x\in A}m(x).
\]

Several notations of msets exist. Among the common notations mentioned
in \cite{ouvrard2018adjacency}, we note in this article a mset $A_{m}$
of universe $A=\left\{ x_{i}:i\in\left\llbracket n\right\rrbracket \right\} $
by: 
\[
A_{m}=\left\{ x_{i}^{m_{i}}:i\in\left\llbracket n\right\rrbracket \right\} 
\]
 where $m_{i}=m\left(x_{i}\right)$.

If $A_{m}$ is a natural multiset, another notation of $A_{m}$ similar
to an unordered list is: 
\[
\left\{ \left\{ \underset{m_{1}\,\text{times}}{\underbrace{x_{1},\ldots,x_{1}}},\ldots,\underset{m_{n}\,\text{times}}{\underbrace{x_{n},\ldots,x_{n}}}\right\} \right\} .
\]

Considering $\mathcal{A}=\Omega_{m_{\mathcal{A}}}$ and $\mathcal{B}=\Omega_{m_{\mathcal{B}}}$
two msets on the same universe $\Omega$, we define the empty mset,
written $\emptyset_{\Omega}$ the set of empty support on the universe
$\Omega$. $\mathcal{A}$ is said to be \textbf{included} in $\mathcal{B}$
- written $\mathcal{A}\subseteq\mathcal{B}$ - if for all $x\in\Omega$:
$m_{\mathcal{A}}(x)\leqslant m_{\mathcal{B}}(x)$. In this case, $\mathcal{A}$
is called a \textbf{submset} of $\mathcal{B}$. The power multiset
of $A$, written $\widetilde{\mathcal{P}}(A)$, is the multiset of
all submsets of $A.$ Different operations can be defined on multisets
of same universe as union, intersection, sum, complementation and
difference: for details one can refer to \cite{singh2007overview}.

\textcolor{blue}{Multisets, under the appellation bag, appear in different
domains such as text modeling, image description and audio \cite{schmitt2016bag}.
In text representation, bag of words have been first introduced in
\cite{harris1954distributional}: bags are lists of words with repetitions,
i.e. multisets of words on a universe. Many applications occur with
different approaches. Bags of words have been used for instance in
fraud detection \cite{purda2015accounting}. More recently bag of
words have been used successfully for translation by neural nets as
a target for the translation as a sentence can be translated in many
different ways \cite{ma2018bag}. In \cite{cummins2018multimodal},
multi-modal bag of words have been used for cross domains sentiment
analysis.}

\textcolor{blue}{Bags of visual words is the transcription to image
of textual bags of words; in bags of visual words, a visual vocabulary
based on image features is built that allows the description of images
as bags of these features. Since their introduction in \cite{sivic2003video},
many applications have been realized: in visual categorization \cite{csurka2004visual},
in image classification and filtering \cite{deselaers2008bag}, in
image annotation \cite{tsai2012bag}, in action recognition \cite{peng2016bag},
in land-use scene classification \cite{zhao2014land}, in identifying
mild traumatic brain injuries \cite{minaee2017identifying} and in
word image retrieval \cite{shekhar2012word}.}

\textcolor{blue}{Bags of concepts are an extension of bags of words
to successive concepts in a text \cite{kim2017bag}. A recent extension
of these concepts is given in \cite{silva2018graph} where bag of
graphs are introduced to encode in graphs the local structure of a
digital object: bags of graphs are declined into bags of singleton
graphs and bags of visual graphs. Using the hb-graphs as we propose
in this article will allow to extend this approach, by taking advantage
of multi-adicity and also of the multiplicity of vertices specific
to each hb-edge.}

\subsection{Hb-graphs}

Hb-graphs are introduced in \cite{ouvrard2018adjacency}. A \textbf{hb-graph}
is a family of multisets with the same universe $V$ and with support
a subset of $V$. The msets are called the \textbf{hb-edges} and the
elements of $V$ the \textbf{vertices}. We consider for the remainder
of the article a hb-graph $\mathcal{H}=\left(V,E\right)$, with $V=\left\{ v_{1},...,v_{n}\right\} $
and $E=\left\{ e_{1},...,e_{p}\right\} $ the family of its hb-edges.

Each hb-edge $e_{i}\in E$ has $V$ as universe and a multiplicity
function associated to it: $m_{e_{i}}:V\rightarrow\mathbb{W}$ where
$\mathbb{W}\subset\mathbb{R}^{+}$. For a general hb-graph, each hb-edge
has to be seen as a weighted system of vertices, where the weights
of each vertex are hb-edge dependent.

A hb-graph where the multiplicity range of each hb-edge is a subset
of $\mathbb{N}$ is called a \textbf{natural hb-graph}. A \textbf{hypergraph}
is a natural hb-graph where the hb-edges have multiplicity one for
every vertex of their support.

The \textbf{order} of a hb-graph $\mathcal{H}$ - written $O\left(\mathcal{H}\right)$
- is: 
\[
O\left(\mathcal{H}\right)=\sum_{v_{i}\in V}\underset{e_{j}\in E}{\max}\left(m_{e_{j}}\left(v_{i}\right)\right).
\]

In a natural hb-graph, the order corresponds to the number of copies
needed to generate the copy hypergraph of the hb-graph.

The \textbf{m-size} of a hb-graph $\mathcal{H}$ - written $s_{m}\left(\mathcal{H}\right)$
- is:

\[
s_{m}\left(\mathcal{H}\right)=\sum_{e_{j}\in E}\sum_{v_{i}\in e_{j}^{\star}}m_{e_{j}}\left(v_{i}\right).
\]

In a natural hb-graph the m-size corresponds to the sum of the m-cardinalities
of the hb-edges of the hb-graph.

The \textbf{support hypergraph} of a hb-graph $\mathcal{H}=\left(V,E\right)$
is the hypergraph whose vertices are the ones of the hb-graph and
whose hyperedges are the support of the hb-edges in a one-to-one way.
We write it $\underline{\mathcal{H}}=\left(V,\underline{E}\right)$,
where $\underline{E}=\left\{ e^{\star}:e\in E\right\} $.

The \textbf{hb-star} of a vertex $v\in V$ is the multiset - written
$H(v)$ and abusively writing $e_{i},\,1\leqslant i\leqslant p$ for
designating the elements of the universe of $H(v)$ corresponding
to the hb-edges of $\mathcal{H}$ of same name - defined as:

\[
H\left(v_{i}\right)=\left\{ e_{j}^{m_{e_{j}}(v_{i})}\,:\,\text{\ensuremath{\forall1\leqslant j\leqslant p}:}\,e_{j}\in E\land v_{i}\in e_{j}^{*}\right\} .
\]

The \textbf{m-degree of a vertex} $v_{i}\in V$ of a hb-graph $\mathcal{H}$
- written $\deg_{m}\left(v_{i}\right)=d_{m}\left(v_{i}\right)$ -
is defined as:

\[
\deg_{m}\left(v_{i}\right)=\#_{m}H\left(v_{i}\right).
\]

We have: 
\[
\sum\limits _{v_{i}\in V}\deg_{m}\left(v_{i}\right)=s_{m}\left(\mathcal{H}\right).
\]

The \textbf{degree of a vertex} $v\in V$ of a hb-graph $\mathcal{H}$
- written $\deg\left(v\right)=d(v)$ - corresponds to the degree of
this vertex in the support hypergraph $\underline{\mathcal{H}}.$

The matrix $H=\left[m_{j}\left(v_{i}\right)\right]_{\substack{1\leqslant i\leqslant n\\
1\leqslant j\leqslant p
}
}$ is called the \textbf{incident matrix} of the hb-graph $\mathcal{H}$.

A \textbf{weighted hb-graph} $\mathcal{H}_{w}=\left(V,E,w_{e}\right)$
is a hb-graph $\mathcal{H}=\left(V,E\right)$ where the hb-edges are
weighted by $w_{e}:E\rightarrow\mathbb{R^{+*}}$. An unweighted hb-graph
is then a weighted hb-graph with $w_{e}\left(e_{j}\right)=1$ for
all $e_{j}\in E$.

A \textbf{strict m-path} $v_{0}e_{1}v_{1}...e_{s}v_{s}$ in a hb-graph
from a vertex $u$ to a vertex $w$ is a vertex / hb-edge alternation
with hb-edges $e_{1}$ to $e_{s}$ and vertices $v_{0}$ to $v_{s}$
such that $v_{0}=u$, $v_{s}=w$, $u\in e_{1}$ and $w\in e_{s}$
and that for all $1\leqslant i\leqslant s-1$, $v_{i}\in e_{i}\cap e_{i+1}$.

A strict m-path $v_{0}e_{1}v_{1}...e_{s}v_{s}$ in a hb-graph corresponds
to a unique path in the hb-graph support hypergraph called the \textbf{support
path}. In this article we abusively call it a path of the hb-graph.
The \textbf{length of a path} corresponds to the number of hb-edges
it is going through.

\begin{figure}
\begin{center}

\begin{tikzpicture}[scale=0.7, every node/.append style={transform shape}]

\node[minimum width=12cm,
		minimum height=1cm,
		anchor=center] (Sentences) {
		};
\node[] (Nmax_Title) at ([yshift=-1.75em,xshift=0cm]Sentences.north) {
	\begin{tabular}{l}
Four sentences:\\
\hspace{0.5cm}\textcolor{S_brique}{$\bullet$ \textbf{P1}: "The sun is in the sky and the sun is yellow."}\\
\hspace{0.5cm}\textcolor{S_green}{$\bullet$ \textbf{P2}: "The sea is blue and the sky is also blue."}\\
\hspace{0.5cm}\textcolor{S_petrol}{$\bullet$ \textbf{P3}: "Navy blue and sky blue are blue colour names."}\\
\hspace{0.5cm}\textcolor{S_purple}{$\bullet$ \textbf{P4}: "Picasso had a blue period where his paintings were in blue shade."}
	\end{tabular}
};

\node[right of=Sentences,
		yshift=-4cm,
		node distance=-5cm,
		minimum width=10cm,
		minimum height=7cm,
		anchor=center] (Graph) {};
\node[ptNode,label=left:sun] (Sun) at ([xshift=-5cm]Graph.north){};
\node[ptNode,label=left:sky] (Sky) at ([xshift=1cm,yshift=-2cm]Sun){};
\node[ptNodeSq,label=left:P2] (P2) at ([xshift=1cm,yshift=-3cm]Sun){};
\node[ptNode,label=left:sea] (Sea) at ([xshift=0cm,yshift=-5cm]Sun){};
\node[ptNodeSq,label=above:P1] (P1) at ([xshift=2cm,yshift=-1cm]Sun){};
\node[ptNode,label=right:yellow] (Yellow) at ([xshift=3cm,yshift=-1cm]Sun){};
\node[ptNodeSq,label=above:P3] (P3) at ([xshift=4cm,yshift=-3cm]Sun){};
\node[ptNode,label=left:blue] (Blue) at ([xshift=3cm,yshift=-4cm]Sun){};
\node[ptNode,label=right:colour] (Colour) at ([xshift=6cm,yshift=-2cm]Sun){};
\node[ptNode,label=right:name] (Name) at ([xshift=6cm,yshift=-3cm]Sun){};
\node[ptNode,label=right:navy] (Navy) at ([xshift=6cm,yshift=-4cm]Sun){};
\node[ptNodeSq,label=below:P4] (P4) at ([xshift=4cm,yshift=-6cm]Sun){};
\node[ptNode,label=right:Picasso] (Picasso) at ([xshift=4cm,yshift=-4cm]Sun){};
\node[ptNode,label=right:period] (Period) at ([xshift=6cm,yshift=-5cm]Sun){};
\node[ptNode,label=right:shade] (Shade) at ([xshift=6cm,yshift=-6cm]Sun){};
\node[ptNode,label=left:painting] (Painting) at ([xshift=3cm,yshift=-7cm]Sun){};
\draw[line width=0.4mm, S_brique] (Sun) -- (P1);
\draw[line width=0.2mm, S_brique] (Sky) -- (P1);
\draw[line width=0.2mm, S_brique] (Yellow) -- (P1);
\draw[line width=0.2mm, S_green] (Sky) -- (P2);
\draw[line width=0.2mm, S_green] (Sea) -- (P2);
\draw[line width=0.2mm, S_green] (Blue) -- (P2);
\draw[line width=0.2mm, S_petrol] (Sky) -- (P3);
\draw[line width=0.6mm, S_petrol] (Blue) -- (P3);
\draw[line width=0.2mm, S_petrol] (Colour) -- (P3);
\draw[line width=0.2mm, S_petrol] (Navy) -- (P3);
\draw[line width=0.2mm, S_petrol] (Name) -- (P3);
\draw[line width=0.2mm, S_purple] (Painting) -- (P4);
\draw[line width=0.2mm, S_purple] (Picasso) -- (P4);
\draw[line width=0.2mm, S_purple] (Period) -- (P4);
\draw[line width=0.4mm, S_purple] (Blue) -- (P4);
\draw[line width=0.2mm, S_purple] (Shade) -- (P4);


\node[right of=Sentences,
		yshift=-4cm,
		node distance=1cm,
		minimum width=6cm,
		minimum height=10cm,
		anchor=center] (Incident) {
\begin{tabular}{|c|c|c|c|c|} \hline   & \textbf{\footnotesize{}\textcolor{S_brique}{P1}} & \textbf{\footnotesize{}\textcolor{S_green}{P2}} & \textbf{\footnotesize{}\textcolor{S_petrol}{P3}} & \textbf{\footnotesize{}\textcolor{S_purple}{P4}}\tabularnewline \hline  {\footnotesize{}sun} & {\footnotesize{}\textcolor{S_brique}{2}} & {\footnotesize{}\textcolor{S_green}{0}} & {\footnotesize{}\textcolor{S_petrol}{0}} & {\footnotesize{}\textcolor{S_purple}{0}}\tabularnewline \hline  {\footnotesize{}sky} & {\footnotesize{}\textcolor{S_brique}{1}} & {\footnotesize{}\textcolor{S_green}{1}} & {\footnotesize{}\textcolor{S_petrol}{1}} & {\footnotesize{}\textcolor{S_purple}{0}}\tabularnewline \hline  {\footnotesize{}yellow} & {\footnotesize{}\textcolor{S_brique}{1}} & {\footnotesize{}\textcolor{S_green}{0}} & {\footnotesize{}\textcolor{S_petrol}{0}} & {\footnotesize{}\textcolor{S_purple}{0}}\tabularnewline \hline  {\footnotesize{}sea} & {\footnotesize{}\textcolor{S_brique}{0}} & {\footnotesize{}\textcolor{S_green}{1}} & {\footnotesize{}\textcolor{S_petrol}{0}} & {\footnotesize{}\textcolor{S_purple}{0}}\tabularnewline \hline  {\footnotesize{}blue} & {\footnotesize{}\textcolor{S_brique}{0}} & {\footnotesize{}\textcolor{S_green}{1}} & {\footnotesize{}\textcolor{S_petrol}{3}} & {\footnotesize{}\textcolor{S_purple}{2}}\tabularnewline \hline  {\footnotesize{}colour} & {\footnotesize{}\textcolor{S_brique}{0}} & {\footnotesize{}\textcolor{S_green}{0}} & {\footnotesize{}\textcolor{S_petrol}{1}} & {\footnotesize{}\textcolor{S_purple}{0}}\tabularnewline \hline  {\footnotesize{}navy} & {\footnotesize{}\textcolor{S_brique}{0}} & {\footnotesize{}\textcolor{S_green}{0}} & {\footnotesize{}\textcolor{S_petrol}{1}} & {\footnotesize{}\textcolor{S_purple}{0}}\tabularnewline \hline  {\footnotesize{}name} & {\footnotesize{}\textcolor{S_brique}{0}} & {\footnotesize{}\textcolor{S_green}{0}} & {\footnotesize{}\textcolor{S_petrol}{1}} & {\footnotesize{}\textcolor{S_purple}{0}}\tabularnewline \hline  {\footnotesize{}painting} & {\footnotesize{}\textcolor{S_brique}{0}} & {\footnotesize{}\textcolor{S_green}{0}} & {\footnotesize{}\textcolor{S_petrol}{0}} & {\footnotesize{}\textcolor{S_purple}{1}}\tabularnewline \hline  {\footnotesize{}Picasso} & {\footnotesize{}\textcolor{S_brique}{0}} & {\footnotesize{}\textcolor{S_green}{0}} & {\footnotesize{}\textcolor{S_petrol}{0}} & {\footnotesize{}\textcolor{S_purple}{1}}\tabularnewline \hline  {\footnotesize{}period} & {\footnotesize{}\textcolor{S_brique}{0}} & {\footnotesize{}\textcolor{S_green}{0}} & {\footnotesize{}\textcolor{S_petrol}{0}} & {\footnotesize{}\textcolor{S_purple}{1}}\tabularnewline \hline  {\footnotesize{}shade} & {\footnotesize{}\textcolor{S_brique}{0}} & {\footnotesize{}\textcolor{S_green}{0}} & {\footnotesize{}\textcolor{S_petrol}{0}} & {\footnotesize{}\textcolor{S_purple}{1}}\tabularnewline \hline  \end{tabular}
};

\end{tikzpicture}

\end{center}

\caption{An example of hb-graphs: four sentences and their associated bag of
words with removed stop words and the incidence matrix of the hb-graph.}
\label{Fig:Hb-graph example}
\end{figure}
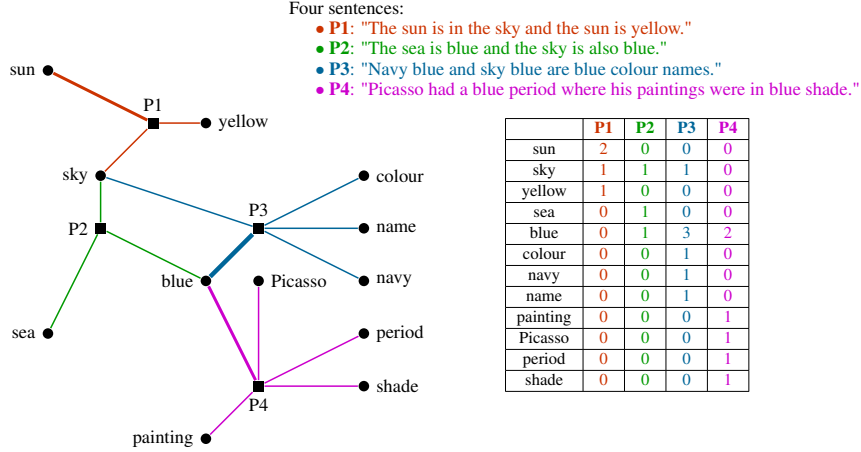

\textcolor{blue}{Representations of hb-graphs can be achieved either
by using sub-mset representations or by using edge representations.
In the edge representation, an extra-node is added per hb-edge and
the thickness of the link between the extra-node of a hb-edge and
the vertices in the support of the hb-edge is made proportional to
the multiplicity of vertices. Except in Figure \ref{Fig:Hb-graph example}
where we use this representation, in this article we use a simplified
representation corresponding to the extra-vertex representation of
the support hypergraph of the hb-graph: an extra-vertex is added for
each hb-edge and the links with the vertices in the support of the
hb-edges are all represented with the same thickness. More details
on these representations can be found in \cite{ouvrard2018adjacency}.}

\textcolor{blue}{We give in Figure \ref{Fig:Hb-graph example} an
example of the representation of a hb-graph of keywords extracted
from sentences in which stop words have been removed. The number of
occurences of the words differs from one sentence to an other: it
is given as a multiplicity that is specific to the corresponding hb-edge
representing the sentence. The universe of the hb-graph is the set
of words where the stop words has been removed.}

\section{Exchange-based diffusion in hb-graphs}

\label{sec:Exchange}

Diffusion processes lead to homogenising information over a structure;
an initial stroke is done on a vertex that propagates over the network
structure. This propagation is often modeled by a random walk on the
network. Random walks in hypergraphs rank vertices by the number of
times they are reached and this ranking is related to the structure
of the network itself. Several random walks with random choices of
the starting vertex are needed to achieve ranking by averaging. Moreover
to avoid loops, teleportation of vertices is needed.

We consider a weighted hb-graph $\mathcal{H}=\left(V,E,w_{e}\right)$
with $\left|V\right|=n$ and $\left|E\right|=p$; we write $H$ the
incident matrix of the hb-graph.

At time $t$ we set a distribution of values over the vertex set:
\[
\alpha_{t}:\left\{ \begin{array}{c}
V\rightarrow\mathbb{R}\\
v_{i}\mapsto\alpha_{t}\left(v_{i}\right)
\end{array}\right..
\]

and a distribution of values over the hb-edge set: 
\[
\epsilon_{t}:\left\{ \begin{array}{c}
E\rightarrow\mathbb{R}\\
e_{j}\mapsto\epsilon_{t}\left(e_{j}\right)
\end{array}\right..
\]

We write $P_{V,t}=\left(\alpha_{t}\left(v_{i}\right)\right)_{1\leqslant i\leqslant n}$
the row state vector of the vertices at time $t$ and $P_{E,t}=\left(\epsilon_{t}\left(e_{j}\right)\right)_{1\leqslant j\leqslant p}$
the row state vector of the hb-edges.

\textcolor{blue}{The initialisation is done such that $\sum\limits _{v_{i}\in V}\alpha_{0}\left(v_{i}\right)=1$
and the information value is concentrated uniformly on the vertices
at the beginning of the diffusion process and, hence, each hb-edge
has a zero value associated to it. Writing $\alpha_{\text{ref}}=\dfrac{1}{\left|V\right|}$,
we set for all $v_{i}\in V:$ $\alpha_{0}\left(v_{i}\right)=\alpha_{\text{ref}}$
and for all $e_{j}\in E$, $\epsilon\left(e_{j}\right)=0.$}

\begin{figure}
\tikzfading [name=arrowfading, top color=transparent!0, bottom color=transparent!95] 
\tikzset{arrowfill/.style={top color=orange!50!white, bottom color=red, general shadow={fill=black, shadow yshift=-0.8ex, path fading=arrowfading}}} 
\tikzset{arrowstyle/.style={draw=orange!50!white,arrowfill, single arrow,minimum height=#1, single arrow, single arrow head extend=.2cm,}}
\begin{center}
\begin{tikzpicture}[scale=0.8, every node/.append style={transform shape}]
\node[text width=0.5cm] at (0.7, -0.5)   (t) {$t$};
\node[text width=1.5cm] at (4.4, -0.5)   (t_demi) {$t+\frac{1}{2}$};
\node[text width=1.5cm] at (7.9, -0.5)   (t_1) {$t+1$};
\node [arrowstyle=1cm,xshift=-0.5cm, yshift=-0.5cm] at ($(t.south)!0.5!(t_demi.south)$) {vertices to hb-edges};
\node [arrowstyle=1cm,xshift=-0.5cm, yshift=-0.5cm] at ($(t_demi.south)!0.5!(t_1.south)$) {hb-edges to vertices};
\node[text width=2cm] at (-1, 2)   (V) {vertices};
\node[state=green] at (0.5, 2) (alpha_t) {\begin{tabular}{c}$v_i$ at \\$\alpha_t\left(v_i\right)$\end{tabular}};
\node[text width=3cm] at (2.75, 1.85) (deps_t) {\begin{tabular}{c}$\delta\epsilon_{t+\frac{1}{2}}\left(e_{j}\mid v_{i}\right)$\end{tabular}};
\node[state=blue] at (4, 1) (eps_t) {\begin{tabular}{c}$e_j$ reaches \\$\epsilon_{t+\frac{1}{2}}\left(e_j\right)$\end{tabular}};
\node[text width=3cm] at (6.75, 1.15) (dalp_t) {\begin{tabular}{c}$\delta\alpha_{t+1}\left(v_{i}\mid e_{j}\right)$\end{tabular}};
\node[state=green] at (7.5, 2) (alpha_t_1) {\begin{tabular}{c}$v_i$ reaches \\$\alpha_{t+1}\left(v_i\right)$\end{tabular}};
\node[text width=2cm] at (-1, 1)   (H) {hb-edges};
\draw[->,ultra thick] (0,0)--(8.1,0) node[right]{}; 
 
\draw[snake=ticks,segment length=2.8cm] (0.5,0) -- (8.1,0);
\path[->,>=stealth'] (alpha_t) edge (eps_t)
(eps_t) edge (alpha_t_1);
\end{tikzpicture}
\end{center}

\caption{Diffusion by exchange: principle}
\label{Fig:Principle_exchange}
\end{figure}
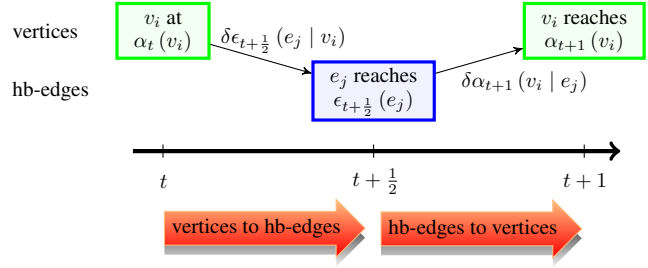

We consider an iterative process with two-phase steps. At every time
step, the first phase starts at time $t$ and ends at $t+\dfrac{1}{2}$
followed by the second phase between time $t+\dfrac{1}{2}$ and $t+1$.
This iterative process is illustrated in Figure \ref{Fig:Principle_exchange}
\textcolor{blue}{that conserves the overall value held by the vertices
and the hb-edges, meaning that we have at any $t\in\left\{ \dfrac{1}{2}k:k\in\mathbb{N}\right\} $:}

\textcolor{blue}{
\[
\sum\limits _{v_{i}\in V}\alpha_{t}\left(v_{i}\right)+\sum\limits _{e_{j}\in E}\epsilon_{t}\left(e_{j}\right)=1.
\]
}

\textbf{During the first phase between time $t$ and $t+\dfrac{1}{2}$},
each vertex $v_{i}$ of the hb-graph shares the value $\alpha_{t}\left(v_{i}\right)$
it holds at time $t$ with the hb-edges it is connected to.

In an unweighted hb-graph, the fraction of $\alpha_{t}\left(v_{i}\right)$
given by $v_{i}$ of m-degree $d_{v_{i}}=\deg_{m}\left(v_{i}\right)$
to each hb-edge is $\dfrac{m_{j}\left(v_{i}\right)}{\deg_{m}\left(v_{i}\right)}$,
which corresponds to the ratio of multiplicity of the vertex $v_{i}$
due to the hb-edge $e_{j}$ over the total $m$-degree of hb-edges
that contains $v_{i}$ in their support.

In a weighted hb-graph, each hb-edge has a weight $w_{e}\left(e_{j}\right)$.
The value $\alpha_{t}\left(v_{i}\right)$ of a vertex $v_{i}$ has
to be shared by taking not only the multiplicity of the vertices in
the hb-edge but also the weight $w_{e}\left(e_{j}\right)$ of a hb-edge
$e_{j}$ into account.

The weights of the hb-edges are stored in a column vector 
\[
w_{E}=\left(w_{e}\left(e_{j}\right)\right)_{1\leqslant j\leqslant p}^{\top}.
\]
We also consider the weight diagonal matrix 
\[
W_{E}=\text{diag}\left(\left(w_{e}\left(e_{j}\right)\right)_{1\leqslant j\leqslant p}\right).
\]

We introduce the weighted $m$-degree matrix: 
\[
D_{w,V}=\text{diag}\left(\left(d_{w,v_{i}}\right)_{1\leqslant i\leqslant n}\right)=\text{diag}\left(Hw_{E}\right).
\]
where $d_{w,v_{i}}$ is called the weighted $m$-degree of the vertex
$v_{i}$. It is: 
\[
d_{w,v_{i}}=\deg_{w,m}\left(v_{i}\right)=\sum\limits _{1\leqslant j\leqslant p}m_{j}\left(v_{i}\right)w_{e}\left(e_{j}\right).
\]
The contribution to the value $\epsilon_{t+\frac{1}{2}}\left(e_{j}\right)$
attached to hb-edge $e_{j}$ of weight $w_{e}\left(e_{j}\right)$
from vertex $v_{i}$ is:
\[
\delta\epsilon_{t+\frac{1}{2}}\left(e_{j}\mid v_{i}\right)=\dfrac{m_{j}\left(v_{i}\right)w_{e}\left(e_{j}\right)}{d_{w,v_{i}}}\alpha_{t}\left(v_{i}\right).
\]
It corresponds to the ratio of weighted multiplicity of the vertex
$v_{i}$ in $e_{j}$ over the total weighted $m$-degree of the hb-edges
where $v_{i}$ is in the support.

We remark that if $v_{i}\notin e_{j}^{\star}$: $\delta\epsilon_{t+\frac{1}{2}}\left(e_{j}\mid v_{i}\right)=0.$

And the value $\epsilon_{t+\frac{1}{2}}\left(e_{j}\right)$ is calculated
by summing over the vertex set:
\[
\epsilon_{t+\frac{1}{2}}\left(e_{j}\right)=\sum\limits _{i=1}^{n}\delta\epsilon_{t+\frac{1}{2}}\left(e_{j}\mid v_{i}\right).
\]

Hence, we obtain: 
\begin{equation}
P_{E,t+\frac{1}{2}}=P_{V,t}D_{w,V}^{-1}HW_{E}\label{eq:P_E_t_0_5}
\end{equation}

\textcolor{blue}{The value given to the hb-edges is subtracted to
the value of the corresponding vertex, hence for all $1\leqslant i\leqslant n$:
\[
\alpha_{t+\frac{1}{2}}\left(v_{i}\right)=\alpha_{t}\left(v_{i}\right)-\sum\limits _{j=1}^{p}\delta\epsilon_{t+\frac{1}{2}}\left(e_{j}\mid v_{i}\right)
\]
}

\begin{claim}[No information on vertices at $t+\dfrac{1}{2}$]

\textcolor{blue}{It holds:}

\textcolor{blue}{
\[
\forall i\in\left\llbracket n\right\rrbracket :\alpha_{t+\frac{1}{2}}\left(v_{i}\right)=0.
\]
}

\end{claim}

\begin{proof}

\textcolor{blue}{For all $i\in\left\llbracket n\right\rrbracket :$
\begin{align*}
\alpha_{t+\frac{1}{2}}\left(v_{i}\right) & =\alpha_{t}\left(v_{i}\right)-\sum\limits _{j=1}^{p}\delta\epsilon_{t+\frac{1}{2}}\left(e_{j}\mid v_{i}\right)\\
 & =\alpha_{t}\left(v_{i}\right)-\sum\limits _{j=1}^{p}\dfrac{m_{j}\left(v_{i}\right)w_{e}\left(e_{j}\right)}{d_{w,v_{i}}}\alpha_{t}\left(v_{i}\right)\\
 & =\alpha_{t}\left(v_{i}\right)-\alpha_{t}\left(v_{i}\right)\dfrac{\sum\limits _{j=1}^{p}m_{j}\left(v_{i}\right)w_{e}\left(e_{j}\right)}{d_{w,v_{i}}}\\
 & =0.
\end{align*}
}

\hfill{}\textcolor{blue}{$\square$}

\end{proof}

\begin{claim}[Conservation of the information of the hb-graph at $t+\dfrac{1}{2}$]

\textcolor{blue}{It holds:}

\textcolor{blue}{
\[
\sum\limits _{v_{i}\in V}\alpha_{t+\frac{1}{2}}\left(v_{i}\right)+\sum\limits _{e\in E}\epsilon_{t+\frac{1}{2}}\left(e\right)=1.
\]
}

\end{claim}

\begin{proof}

\textcolor{blue}{We have:}

\textcolor{blue}{
\begin{align*}
\sum\limits _{v_{i}\in V}\alpha_{t+\frac{1}{2}}\left(v_{i}\right)+\sum\limits _{e\in E}\epsilon_{t+\frac{1}{2}}\left(e\right) & =\sum\limits _{e_{j}\in E}\epsilon_{t+\frac{1}{2}}\left(e_{j}\right)\\
 & =\sum\limits _{e_{j}\in E}\sum\limits _{i=1}^{n}\delta\epsilon_{t+\frac{1}{2}}\left(e_{j}\mid v_{i}\right)\\
 & =\sum\limits _{e_{j}\in E}\sum\limits _{i=1}^{n}\dfrac{m_{j}\left(v_{i}\right)w_{e}\left(e_{j}\right)}{d_{w,v_{i}}}\alpha_{t}\left(v_{i}\right)\\
 & =\sum\limits _{i=1}^{n}\alpha_{t}\left(v_{i}\right)\dfrac{\sum\limits _{e_{j}\in E}m_{j}\left(v_{i}\right)w_{e}\left(e_{j}\right)}{d_{w,v_{i}}}\\
 & =\sum\limits _{i=1}^{n}\alpha_{t}\left(v_{i}\right)\\
 & =1
\end{align*}
}

\textcolor{blue}{\hfill{}$\square$}

\end{proof}

\textbf{During the second phase that starts at time $t+\dfrac{1}{2}$},
the hb-edges share their values across the vertices they hold taking
into account the multiplicity of the vertices in the hb-edge. Every
value is modulated by the weight $w_{e}\left(e_{j}\right)$ of the
hb-edge $e_{j}$ it comes from.

The contribution to $\alpha_{t+1}\left(v_{i}\right)$ given by a hb-edge
$e_{j}$ is propotional to $\epsilon_{t+\frac{1}{2}}$ in a factor
corresponding to the ratio of the multiplicity $m_{j}\left(v_{i}\right)$
of the vertex $v_{i}$ to the hb-edge m-cardinality:

\textcolor{blue}{
\[
\delta\alpha_{t+1}\left(v_{i}\mid e_{j}\right)=\dfrac{m_{j}\left(v_{i}\right)}{\#_{m}e_{j}}\epsilon_{t+\frac{1}{2}}\left(e_{j}\right).
\]
}

The value $\alpha_{t+1}\left(v_{i}\right)$ is then obtained by summing
on all values associated to the hb-edges that are incident to $v_{i}$:

\[
\alpha_{t+1}\left(v_{i}\right)=\sum\limits _{j=1}^{p}\delta\alpha_{t+1}\left(v_{i}\mid e_{j}\right).
\]

Writing $D_{E}=\text{diag}\left(\#_{m}e_{j}\right)_{1\leqslant j\leqslant p}$
the diagonal matrix of size $p\times p$, it comes: 
\begin{equation}
P_{E,t+\frac{1}{2}}D_{E}^{-1}H^{\top}=P_{V,t+1}.\label{eq:P_V_t_1}
\end{equation}

\textcolor{blue}{The values given to the vertices are subtracted to
the value associated to the corresponding hb-edge. Hence, for all
$1\leqslant j\leqslant p$: 
\[
\epsilon_{t+1}\left(e_{j}\right)=\epsilon_{t+\frac{1}{2}}\left(e_{j}\right)-\sum\limits _{i=1}^{n}\delta\alpha_{t+1}\left(v_{i}\mid e_{j}\right)
\]
}

\begin{claim}[The hb-edges have no value at $t+1$]

\textcolor{blue}{It holds:}

\textcolor{blue}{
\[
\epsilon_{t+1}\left(e_{j}\right)=0.
\]
}

\end{claim}

\begin{proof}

\textcolor{blue}{For all $i\in\left\llbracket p\right\rrbracket :$}

\textcolor{blue}{
\begin{align*}
\epsilon_{t+1}\left(e_{j}\right) & =\epsilon_{t+\frac{1}{2}}\left(e_{j}\right)-\sum\limits _{i=1}^{n}\delta\alpha_{t+1}\left(v_{i}\mid e_{j}\right)\\
 & =\epsilon_{t+\frac{1}{2}}\left(e_{j}\right)-\sum\limits _{i=1}^{n}\dfrac{m_{j}\left(v_{i}\right)}{\#_{m}e_{j}}\epsilon_{t+\frac{1}{2}}\left(e_{j}\right)\\
 & =\epsilon_{t+\frac{1}{2}}\left(e_{j}\right)\left(1-\dfrac{\sum\limits _{i=1}^{n}m_{j}\left(v_{i}\right)}{\#_{m}e_{j}}\right)\\
 & =0.
\end{align*}
}

\textcolor{blue}{\hfill{}$\square$}

\end{proof}

\begin{claim}[Conservation of the information of the hb-graph at $t+1$]

\textcolor{blue}{It holds:}

\textcolor{blue}{
\[
\sum\limits _{v_{i}\in V}\alpha_{t+1}\left(v_{i}\right)+\sum\limits _{e_{j}\in E}\epsilon_{t+1}\left(e_{j}\right)=1.
\]
}

\end{claim}

\begin{proof}

\textcolor{blue}{
\begin{align*}
\sum\limits _{v_{i}\in V}\alpha_{t+1}\left(v_{i}\right)+\sum\limits _{e\in E}\epsilon_{t+1}\left(e\right) & =\sum\limits _{v_{i}\in V}\alpha_{t+1}\left(v_{i}\right)\\
 & =\sum\limits _{v_{i}\in V}\sum\limits _{j=1}^{p}\delta\alpha_{t+1}\left(v_{i}\mid e_{j}\right)\\
 & =\sum\limits _{v_{i}\in V}\sum\limits _{j=1}^{p}\dfrac{m_{j}\left(v_{i}\right)}{\#_{m}e_{j}}\epsilon_{t+\frac{1}{2}}\left(e_{j}\right)\\
 & =\sum\limits _{j=1}^{p}\epsilon_{t+\frac{1}{2}}\left(e_{j}\right)\dfrac{\sum\limits _{v_{i}\in V}m_{j}\left(v_{i}\right)}{\#_{m}e_{j}}\\
 & =\sum\limits _{j=1}^{p}\epsilon_{t+\frac{1}{2}}\left(e_{j}\right)\\
 & =1.
\end{align*}
}

\textcolor{blue}{\hfill{}$\square$}

\end{proof}

Regrouping (\ref{eq:P_E_t_0_5}) and (\ref{eq:P_V_t_1}): 
\begin{equation}
P_{V,t+1}=P_{V,t}D_{w,V}^{-1}HW_{E}D_{E}^{-1}H^{\top}.\label{eq:P_V_tplus1}
\end{equation}
It is valuable to keep a trace of the intermediate state $P_{E,t+\frac{1}{2}}=P_{V,t}D_{w,V}^{-1}HW_{E}$
as it records the importance of the hb-edges.

Writing $T=D_{w,V}^{-1}HW_{E}D_{E}^{-1}H^{\top}$, it follows from
\ref{eq:P_V_tplus1}:
\begin{equation}
P_{V,t+1}=P_{V,t}T.\label{eq:transition_state}
\end{equation}

\begin{claim}[Stochastic transition matrix]

\textcolor{blue}{$T$ is a square row stochastic matrix of dimension
$n.$}

\end{claim}

\begin{proof}

\textcolor{blue}{Let consider: $A=\left(a_{ij}\right)_{\substack{1\leqslant i\leqslant n\\
1\leqslant j\leqslant p
}
}=D_{w,V}^{-1}HW_{E}\in M_{n,p}$ and $B=\left(b_{jk}\right)_{\substack{1\leqslant j\leqslant p\\
1\leqslant k\leqslant n
}
}=D_{E}^{-1}H^{\top}\in M_{p,n}.$}

\textcolor{blue}{$A$ and $B$ are nonnegative rectangular matrices.
Moreover:}

\textcolor{blue}{$a_{ij}=\dfrac{m_{j}\left(v_{i}\right)w_{e}\left(e_{j}\right)}{d_{w,v_{i}}}$
and it holds: 
\[
\sum\limits _{j=1}^{p}a_{ij}=\dfrac{\sum\limits _{j=1}^{p}m_{j}\left(v_{i}\right)w_{e}\left(e_{j}\right)}{d_{w,v_{i}}}=1.
\]
}

\textcolor{blue}{$b_{jk}=\dfrac{m_{j}\left(v_{k}\right)}{\#_{m}\left(e_{j}\right)}$
and it holds:
\[
\sum\limits _{k=1}^{n}b_{jk}=\dfrac{\sum\limits _{k=1}^{n}m_{j}\left(v_{k}\right)}{\#_{m}e_{j}}=1.
\]
}

\textcolor{blue}{We have: $P_{V,t+1}=P_{V,t}AB$ where:}

\textcolor{blue}{
\[
AB=\left(\sum\limits _{j=1}^{p}a_{ij}b_{jk}\right)_{\substack{1\leqslant i\leqslant n\\
1\leqslant k\leqslant n
}
}.
\]
}

\textcolor{blue}{It yields:}

\textcolor{blue}{
\begin{align*}
\sum\limits _{k=1}^{n}\sum\limits _{j=1}^{p}a_{ij}b_{jk} & =\sum\limits _{j=1}^{p}a_{ij}\sum\limits _{k=1}^{n}b_{jk}\\
 & =\sum\limits _{j=1}^{p}a_{ij}\\
 & =1.
\end{align*}
}

\textcolor{blue}{Hence $AB$ is a nonnegative square matrix with its
row sums all equal to 1: it is a row stochastic matrix.}

\textcolor{blue}{\hfill{}$\square$}

\end{proof}

\begin{claim}[Properties of T]

\textcolor{blue}{Supposing that the hb-graph is connected, the exchange-based
diffusion matrix $T$ is aperiodic and irreducible.}

\end{claim}

\begin{proof}

\textcolor{blue}{This stochastic matrix is aperiodic, due to the fact
that any vertex of the hb-graph retrieves a part of the value it has
given to the hb-edge, hence $t_{ii}>0$ for all $1\leqslant i\leqslant n$.}

\textcolor{blue}{Moreover as the hb-graph is connected, the matrix
is irreducible as all state can be joined from any state.}

\textcolor{blue}{\hfill{}$\square$}

\end{proof}

\begin{claim}

The sequence $\left(P_{V,t}\right)_{t\in\mathbb{N}}$, with $P_{V,t}=\left(\alpha_{t}\left(v_{i}\right)\right)_{1\leqslant i\leqslant n}$
in a connected hb-graph converges to the state vector $\pi_{V}$ such
that: 
\[
\pi_{V}=\left(\dfrac{d_{w,v_{i}}}{\sum\limits _{k=1}^{n}d_{w,v_{k}}}\right)_{1\leqslant i\leqslant n}.
\]

\end{claim}

\begin{proof}

\textcolor{blue}{We designate by $\pi$ an eigenvector of $T$ associated
to the eigenvalue 1. We have $\pi T=\pi.$}

\textcolor{blue}{Let consider $u=\left(d_{w,v_{i}}\right)_{1\leqslant i\leqslant n}.$}

\textcolor{blue}{We have 
\begin{align*}
\left(uT\right)_{k} & =\sum\limits _{i=1}^{n}d_{w,v_{i}}\sum\limits _{j=1}^{p}c_{ik}\\
 & =\sum\limits _{i=1}^{n}d_{w,v_{i}}\sum\limits _{j=1}^{p}\dfrac{m_{j}\left(v_{i}\right)w_{e}\left(e_{j}\right)}{d_{w,v_{i}}}\times\dfrac{m_{j}\left(v_{k}\right)}{\#_{m}\left(e_{j}\right)}\\
 & =\sum\limits _{j=1}^{p}\sum\limits _{i=1}^{n}m_{j}\left(v_{i}\right)w_{e}\left(e_{j}\right)\times\dfrac{m_{j}\left(v_{k}\right)}{\#_{m}\left(e_{j}\right)}\\
 & =\sum\limits _{j=1}^{p}w_{e}\left(e_{j}\right)m_{j}\left(v_{k}\right)\dfrac{\sum\limits _{i=1}^{n}m_{j}\left(v_{i}\right)}{\#_{m}\left(e_{j}\right)}\\
 & =\sum\limits _{j=1}^{p}w_{e}\left(e_{j}\right)m_{j}\left(v_{k}\right)\\
 & =d_{w,v_{k}}=u_{k}
\end{align*}
}

\textcolor{blue}{Hence, $u$ is a nonnegative eigenvector of $T$
associated to the eigenvalue 1.}

\textcolor{blue}{When we iterate over $T$ which is a stochastic matrix
aperiodic and irreducible for a connected hb-graph we are then ensured
to converge to a stationary state which is the probability vector
associated to the eigenvalue 1. It is unique and is equal to $\alpha u$
such that $\sum\limits _{k=1}^{n}\alpha u_{k}=1$.}

\textcolor{blue}{We have $\alpha=\dfrac{1}{\sum\limits _{k=1}^{n}d_{w,v_{k}}}$
and hence the result.}

\end{proof}

\begin{claim}

\textcolor{blue}{The sequence $\left(P_{E,t+\frac{1}{2}}\right)_{t\in\mathbb{N}}$,
with $P_{E,t+\frac{1}{2}}=\left(\epsilon_{t+\frac{1}{2}}\left(e_{j}\right)\right)_{1\leqslant j\leqslant p}$
in a connected hb-graph converges to the state vector $\pi_{E}$ such
that: $\left(\dfrac{w_{e}\left(e_{j}\right)\times\#_{m}\left(e_{j}\right)}{\sum\limits _{k=1}^{n}d_{w,v_{k}}}\right)_{1\leqslant j\leqslant p}.$}

\end{claim}

\begin{proof}

\textcolor{blue}{As $P_{E,t+\frac{1}{2}}=P_{V,t}D_{w,V}^{-1}HW_{E}$
and that $\lim\limits _{t\to+\infty}P_{V,t}=\pi_{V}$, the sequence
$\left(P_{E,t+\frac{1}{2}}\right)_{t\in\mathbb{N}}$ converges towards
a state vector $\pi_{E}$ such that: $\pi_{E}=\pi_{V}D_{w,V}^{-1}HW_{E}.$}

\textcolor{blue}{We have: 
\begin{align*}
\pi_{E} & =\left(\sum\limits _{i=1}^{n}\dfrac{d_{w,v_{i}}}{\sum\limits _{k=1}^{n}d_{w,v_{k}}}\times\dfrac{m_{j}\left(v_{i}\right)\times w_{e}\left(e_{j}\right)}{d_{w,v_{i}}}\right)_{1\leqslant j\leqslant p}\\
 & =\left(\sum\limits _{i=1}^{n}\dfrac{m_{j}\left(v_{i}\right)\times w_{e}\left(e_{j}\right)}{\sum\limits _{k=1}^{n}d_{w,v_{k}}}\right)_{1\leqslant j\leqslant p}\\
 & =\left(\dfrac{w_{e}\left(e_{j}\right)\times\sum\limits _{i=1}^{n}m_{j}\left(v_{i}\right)}{\sum\limits _{k=1}^{n}d_{w,v_{k}}}\right)_{1\leqslant j\leqslant p}\\
 & =\left(\dfrac{w_{e}\left(e_{j}\right)\times\#_{m}\left(e_{j}\right)}{\sum\limits _{k=1}^{n}d_{w,v_{k}}}\right)_{1\leqslant j\leqslant p}.
\end{align*}
}

\textcolor{blue}{All components are nonnegative and we check that
the components of this vector sum to one: 
\begin{align*}
\sum\limits _{j=1}^{p}\pi_{E,j} & =\dfrac{\sum\limits _{j=1}^{p}w_{e}\left(e_{j}\right)\times\sum\limits _{i=1}^{n}m_{j}\left(v_{i}\right)}{\sum\limits _{k=1}^{n}d_{w,v_{k}}}\\
 & =\dfrac{\sum\limits _{i=1}^{n}\sum\limits _{j=1}^{p}w_{e}\left(e_{j}\right)\times m_{j}\left(v_{i}\right)}{\sum\limits _{k=1}^{n}d_{w,v_{k}}}\\
 & =\dfrac{\sum\limits _{i=1}^{n}d_{w,v_{i}}}{\sum\limits _{k=1}^{n}d_{w,v_{k}}}\\
 & =1.
\end{align*}
}

\end{proof}

\textcolor{blue}{These two claims show that this exchange-based process
ranks vertices by their weighted m-degree and of hb-edges by their
weighted m-cardinality.}

\textcolor{blue}{We have gathered the two-phase steps of the exchange-based
diffusion process in Algorithm \ref{Alg: Exchange}. The time complexity
of this algorithm is $O\left(T\text{\ensuremath{\left(d_{\mathcal{H}}n+r_{\mathcal{H}}p\right)}}\right)$
where $d_{\mathcal{H}}=\underset{v_{i}\in V}{\max}\left(d_{i}\right)$
is the maximal degree of vertices in the hb-graph and $r_{\mathcal{H}}=\underset{e_{j}\in E}{\max}\left|e_{j}^{\star}\right|$
is the maximal cardinality of the support of a hb-graph. Usually,
$d_{\mathcal{H}}$ and $r_{\mathcal{H}}$ are small compared to $n$
and $p.$ Algorithm \ref{Alg: Exchange} can be refined to determine
automatically the number of iterations needed by fixing an accepted
error to ensure convergence on the values of the vertices and storing
the previous state.}

\begin{algorithm}
\textbf{\textcolor{blue}{Given:}}

\textcolor{blue}{\hspace{0.5cm}A hb-graph $\mathcal{H}=\left(V,E,w_{e}\right)$
with $\left|V\right|=n$ and $\left|E\right|=p$}

\textcolor{blue}{\hspace{0.5cm}Number of iterations: $T$}\\

\textbf{\textcolor{blue}{Initialisation:}}

\textcolor{blue}{\hspace{0.5cm}For all $v_{i}\in V:$ $\alpha_{i}:=\dfrac{1}{n}$}

\textcolor{blue}{\hspace{0.5cm}For all $e_{j}\in E:$ $\epsilon_{j}:=0$}\\

\textbf{\textcolor{blue}{DiffuseFromVerticesToHbEdges():}}

\textcolor{blue}{\hspace{0.5cm}For $j:=1$ to $p$:}

\textcolor{blue}{\hspace{0.5cm}\hspace{0.5cm}$\epsilon_{j}:=0$}

\textcolor{blue}{\hspace{0.5cm}\hspace{0.5cm}For $v_{i}\in e_{j}^{\star}$:}

\textcolor{blue}{\hspace{0.5cm}\hspace{0.5cm}\hspace{0.5cm}$\epsilon_{j}:=\epsilon_{j}+\dfrac{m_{j}\left(v_{i}\right)w_{e}\left(e_{j}\right)}{d_{w,m}\left(v_{i}\right)}\alpha_{i}$}\\

\textbf{\textcolor{blue}{DiffuseFromHbEdgesToVertices():}}

\textcolor{blue}{\hspace{0.5cm}For $i:=1$ to $n$:}

\textcolor{blue}{\hspace{0.5cm}\hspace{0.5cm}$\alpha_{i}:=0$}

\textcolor{blue}{\hspace{0.5cm}\hspace{0.5cm}For $e_{j}$ such that
$v_{i}\in e_{j}^{\star}$:}

\textcolor{blue}{\hspace{0.5cm}\hspace{0.5cm}\hspace{0.5cm}$\alpha_{i}:=\alpha_{i}+\dfrac{m_{j}\left(v_{i}\right)}{\#_{m}e_{j}}\epsilon_{j}$}\\

\textbf{\textcolor{blue}{Main():}}

\textcolor{blue}{\hspace{0.5cm}Calculate for all $i:$ $d_{w,m}\left(v_{i}\right)$
and for all $j:$ $\#_{m}e_{j}$}

\textcolor{blue}{\hspace{0.5cm}For $t=1$ to $T$:}

\textcolor{blue}{\hspace{0.5cm}\hspace{0.5cm}DiffuseFromVerticesToHbEdges()}

\textcolor{blue}{\hspace{0.5cm}\hspace{0.5cm}DiffuseFromHbEdgesToVertices()}

\caption{Exchange-based diffusion}

\label{Alg: Exchange}
\end{algorithm}

\section{Results and evaluation}

\label{sec:Results}

\textcolor{blue}{This section firstly adresses the validation of the
approach taken on random hb-graphs. Secondly, this approach is applied
to help in the processing of the results of Arxiv querying.}

\subsection{Validation on random hb-graphs}

This diffusion by exchange process has been validated on two experiments:
the first experiment generates a random hb-graph to validate our approach
and the second compares the results to a classical random walk on
the hb-graph.

\begin{figure}
\begin{center}

\begin{tikzpicture}[->,>=stealth',scale=0.85, every node/.append style={transform shape}]

\node[state=purple,
		yshift=-1cm,
		minimum width=12cm,
		minimum height=1cm,
		anchor=center] (Nmax) {
		};
\node[] (Nmax_Title) at ([yshift=-1.75em,xshift=0cm]Nmax.north) {\textbf{$N_\textrm{max}$ vertices are generated}};

\node[state=red,
		right of=Nmax,
		yshift=-2cm,
		node distance=-4cm,
		minimum width=4cm,
		minimum height=1cm,
		anchor=center] (N0) {$V_0$};
\node[state=green!50!black,
		right of=Nmax,
		yshift=-2cm,
		node distance=3.5cm,
		minimum width=1cm,
		minimum height=1cm,
		dashed,
		anchor=center] (N0_3_1) {};
\node[state=green!50!black,
		right of=Nmax,
		yshift=-2cm,
		node distance=0.5cm,
		minimum width=1cm,
		minimum height=1cm,
		dashed,
		anchor=center] (N0_3_2) {};

\node[state=green!50!black,
		right of=Nmax,
		yshift=-2cm,
		node distance=-0.9cm,
		minimum width=2cm,
		minimum height=1cm,
		anchor=center] (N0_0) {$V_1$};
\node[state=green!50!black,
		right of=Nmax,
		yshift=-2cm,
		node distance=2cm,
		minimum width=2cm,
		minimum height=1cm,
		anchor=center] (N0_1) {$V_j$};

\node[state=green!50!black,
		right of=Nmax,
		yshift=-2cm,
		node distance=5cm,
		minimum width=2cm,
		minimum height=1cm,
		anchor=center] (N0_5) {$V_j$};
\node[state=green!50!black,
		right of=Nmax,
		yshift=-2cm,
		node distance=5cm,
		minimum width=2cm,
		minimum height=1cm,
		anchor=center] (N0_7) {$V_k$};

\node[state=orange,
		right of=N0_1,
		yshift=-2cm,
		node distance=-3cm,
		minimum width=3cm,
		minimum height=1cm,
		anchor=center] (N0_detail_left) {\begin{tabular}{c}$N_{j,1}$ \\ important vertices\end{tabular}};
\node[state=orange,
		right of=N0_1,
		yshift=-2cm,
		node distance=1.5cm,
		minimum width=6cm,
		minimum height=1cm,
		anchor=center] (N0_detail_right) {\begin{tabular}{c}$N_{j,2}$ \\ remaining vertices\end{tabular}};
\fill[fill=red!10!white] ([xshift=-2cm,yshift=1pt]N0.north)--([yshift=-1pt]Nmax.south)--([xshift=2cm,yshift=1pt]N0.north);
\fill[fill=green!50!black!10!white] ([xshift=2cm,yshift=1pt]N0.north)--([yshift=-1pt]Nmax.south)--([xshift=1cm,yshift=1pt]N0_7.north);
\node[] (N0_Title) at ([yshift=0.5em,xshift=0cm]N0.north) {\textcolor{red}{\textbf{$N_0$ interconnected vertices}}};

\node[] (N0_Title) at ([yshift=0.5em,xshift=3cm]N0_0.north) {\textcolor{green!50!black}{\textbf{$N_\textrm{max}-N_0$ vertices in $k$ groups}}};

\fill[fill=green!50!black!10!white] ([xshift=-1.5cm,yshift=1pt]N0_detail_left.north)--([yshift=-1pt]N0_1.south)--([xshift=3cm,yshift=1pt]N0_detail_right.north);

\node[] (N0_detail_Title) at ([yshift=-2em,xshift=0cm]N0_1.south) {$N_{j,1}\ll N_{j,2}$};

\end{tikzpicture}

\end{center}

\caption{Random hb-graph generation principle}

\label{Fig: Hb-graph generation principle}
\end{figure}
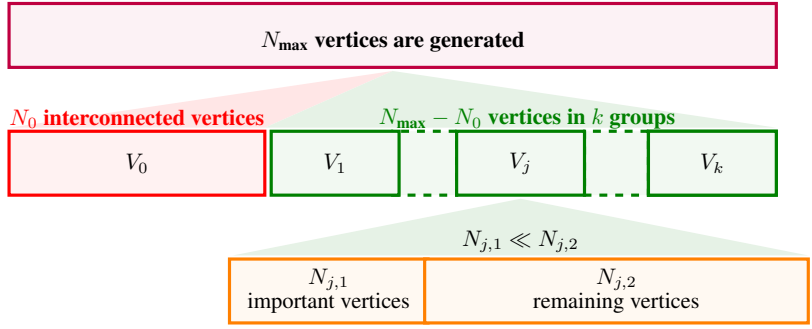

We built a random unweighted hb-graph generator. The generator makes
it possible to construct a hb-graph with inter-connected sub-hb-graphs;
those sub-hb-graphs can be potentially disconnected leading to multiple
connected components. We restricted ourselves in the experiments to
connected hb-graphs. A single connected component is built by choosing
the number of intermediate vertices that link the different sub-hb-graphs
together. As it is show in Figure \ref{Fig: Hb-graph generation principle},
we generate $N_{\text{max}}$ vertices. We start by building each
sub-hb-graph, called group, individually and then interconnect them.
Let $k$ be the number of groups. A first set $V_{0}$ of interconnected
vertices is built by choosing $N_{0}$ vertices out of the $N_{\text{max}}$.
The remaining $N_{\text{max}}-N_{0}$ vertices are then separated
into $k$ subsets $\left(V_{j}\right)_{1\leqslant j\leqslant k}$.
In each of these $k$ groups $V_{j}$ we generate two subsets of vertices:
a first set $V_{j,1}$ of $N_{j,1}$ vertices and a second set $V_{j,2}$
of $N_{j,2}$ vertices with $N_{j,1}\ll N_{j,2}$, $1\leqslant j\leqslant k$.
The number of hb-edges to be built is adjustable: their number is
shared between the different groups. The m-cardinality $\#_{m}\left(e\right)$
of a hb-edge is chosen randomly below a maximum tunable threshold.
The $V_{j,1}$-vertices are considered as important vertices and must
be present in a certain number of hb-edges per group; the number of
important vertices in a hb-edge is randomly fixed below a maximum
number. The completion of the hb-edge is done by choosing vertices
randomly in the $V_{j,2}$ set. The random choice made into these
two groups is tuned to follow a power law distribution. It implies
that some vertices occur more often than others. Interconnection between
the $k$ components is achieved by choosing vertices in $V_{0}$ and
inserting them randomly into the hb-edges built.

We apply the exchange-based diffusion process on these generated hb-graphs:
after a few iterations, we visualize the hb-graphs to observe the
evolution of the vertex values using a gradient coloring scale. We
also take advantage of the first half-step to highlight hb-edges in
the background and show hb-edge importance using an other gradient
coloring scale.

To get proper evaluation and show that vertices with the highest $\alpha$-values
correspond to the important vertices of the network - in the sense
of being central for the connectivity - we compute the eccentricity
of vertices from a subset $S$ of the vertex set $V$ to the remaining
$V\backslash S$ of the vertices. The eccentricity of a vertex in
a graph is the length of a maximal shortest path between this vertex
and the other vertices of this graph: extending this definition to
hb-graphs is straightforward. If the graph is disconnected then each
vertex has infinite eccentricity.

\begin{figure}
\begin{center}

\begin{tikzpicture}[scale=1.7]

\path
coordinate (A) at (3,0)   
coordinate (B) at (0,1.45)   
coordinate (C) at (-3,0)   
coordinate (D) at (0,-1.45)
coordinate (E) at (-1.5,1.29904)
coordinate (F) at (-1.5,-1.29904)
;
\draw[S_green, ultra thick]
	(A) to [out=90,in=0]
	(B) to [out=180, in=90]
	(C) to [out=270, in=180]
	(D) to [out=0, in=270]
	cycle;
\draw[S_green, ultra thick] (E) to [out=-60,in=60] (F); 
\node[S_green, ultra thick] (V) at (3.2,1) {\textbf{V}};
\node[S_petrol, ultra thick] (A) at (-2,0) {$\bold{S}$};
\node[S_petrol, ultra thick] (B) at (1,0) {$\bold{V \backslash S}$};

\node[] (AccrG) at (-2,1){};
\node[ptNode, label=left:$v_0$, S_brique] (G1) at ([xshift=0.01cm]AccrG){};
\node[ptNode] (G2) at ([xshift=1cm]AccrG){};
\node[ptNode] (G3) at ([xshift=0.5cm,yshift=-1cm]AccrG){};
\node[ptNode] (G4) at ([xshift=1cm,yshift=-1cm]AccrG){};
\node[ptNode] (G5) at ([xshift=2cm,yshift=0.25cm]AccrG){};
\node[ptNode] (G6) at ([xshift=3.5cm,yshift=-0.25cm]AccrG){};
\node[ptNode] (G7) at ([xshift=4cm,yshift=-1cm]AccrG){};
\node[ptNode] (G8) at ([xshift=0.5cm,yshift=-1cm]AccrG){};
\node[ptNode] (G9) at ([xshift=2cm,yshift=-2cm]AccrG){};
\node[ptNode] (G10) at ([xshift=2cm,yshift=-1.5cm]AccrG){};
\node[] (G11) at ([xshift=2.5cm,yshift=-1.5cm]AccrG){};
\node[] (G12) at ([xshift=2.5cm,yshift=-1cm]AccrG){};
\node[] (G13) at ([xshift=2.5cm,yshift=-2cm]AccrG){};
\node[] (G14) at ([xshift=1.7cm,yshift=-0.2cm]AccrG){};
\node[] (G15) at ([xshift=2cm,yshift=-0.5cm]AccrG){};
\node[] (G16) at ([xshift=2cm,yshift=-1.5cm]AccrG){};
\node[] (G17) at ([xshift=2.5cm,yshift=-0.5cm]AccrG){};
\node[] (G18) at ([xshift=2.7cm,yshift=-0.8cm]AccrG){};
\node[] (G19) at ([xshift=2.75cm,yshift=-2cm]AccrG){};
\node[] (G20) at ([xshift=2.75cm,yshift=-1.5cm]AccrG){};
\node[ptNodeSq] (P1) at ([xshift=0.6cm,yshift=-0.5cm]AccrG){};
\node[ptNodeSq] (P2) at ([xshift=1.3cm,yshift=-1.5cm]AccrG){};
\node[ptNodeSq] (P5) at ([xshift=1.5cm,yshift=-0.5cm]AccrG){};
\node[ptNodeSq] (P3) at ([xshift=3cm,yshift=-0.5cm]AccrG){};
\node[ptNodeSq] (P4) at ([xshift=3.25cm,yshift=-1.5cm]AccrG){};

\draw[line width=0.4mm, S_brique] (G1) -- (P1);
\draw[line width=0.6mm] (G2) -- (P1);
\draw[line width=0.2mm] (G3) -- (P1);
\draw[line width=0.8mm, S_brique] (G4) -- (P1);
\draw[line width=0.6mm, dashed, S_brique] (G4) -- (P2);
\draw[line width=0.4mm] (G3) -- (P2);
\draw[line width=0.2mm] (G9) -- (P2);
\draw[line width=0.4mm,dashed,S_brique] (G10) -- (P2);
\draw[line width=0.4mm,dashed,S_brique] (G10) -- (G11);
\draw[line width=0.8mm,dashed] (G10) -- (G12);
\draw[line width=0.4mm,dashed] (G10) -- (G13);
\draw[line width=0.6mm,dashed] (P5) -- (G14);
\draw[line width=0.4mm,dashed] (P5) -- (G15);
\draw[line width=0.2mm,dashed] (P5) -- (G16);
\draw[line width=0.4mm,dashed] (G4) -- (P5);
\draw[line width=0.4mm] (G5) -- (P3);
\draw[line width=0.2mm] (G6) -- (P3);
\draw[line width=0.6mm] (G7) -- (P3);
\draw[line width=0.8mm,S_brique] (G6) -- (P4);
\draw[line width=0.8mm,S_brique] (G7) -- (P4);

\draw[line width=0.2mm,dashed] (G17) -- (P3);
\draw[line width=0.6mm,dashed] (G18) -- (P3);
\draw[line width=0.8mm,dashed] (G19) -- (P4);
\draw[line width=0.8mm,dashed,S_brique] (G20) -- (P4);

\end{tikzpicture}

\end{center}

\caption{Relative excentricity: finding the length of a maximal shortest path
in the hb-graph starting from a given vertex $v_{0}$ of $S$ and
finishing with any vertex in $V\backslash S$}

\label{Fig: Hb-graph relative excentricity}
\end{figure}
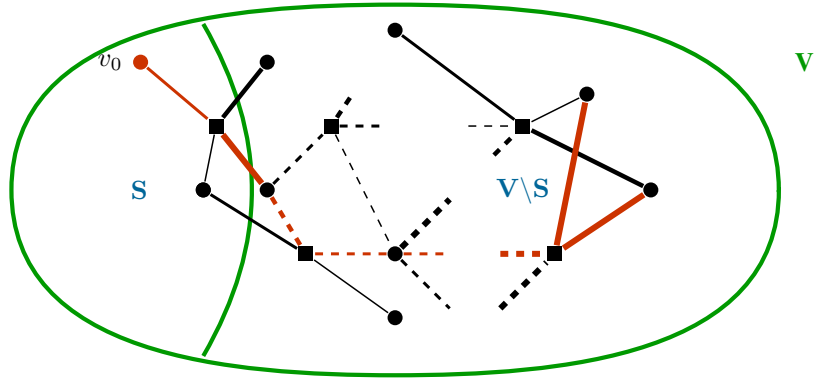

For the purpose of evaluation, in this article, we define a \textbf{relative
eccentricity} as the length of a maximal shortest path starting from
a given vertex in $S$ and ending with any vertices of $V\backslash S$;
the relative eccentricity is calculated for each vertex of $S$ provided
that it is connected to vertices of $V\backslash S$; otherwise it
is set to $-\infty$. The concept of relative excentricity is illustrated
in Figure \ref{Fig: Hb-graph relative excentricity}.

For the vertex set $V$, the subset used for relative eccentricity
is built by using a threshold value $s_{V}$: vertices with $\alpha$
value above this threshold are gathered into a subset $A_{V}\left(s_{V}\right)$
of $V$. We consider $B_{V}\left(s_{V}\right)=V\backslash A_{V}\left(s_{V}\right)$,
the set of vertices with $\alpha$ values below this threshold. We
evaluate the relative eccentricity of each vertex of $A_{V}\left(s_{V}\right)$
to vertices of $B_{V}\left(s_{V}\right)$ in the support hypergraph
of the corresponding hb-graph.

Assuming that we stop iterating at time $T$, we let $s_{V}$ vary
from 0 to the value $\alpha_{T,\max}=\underset{v\in V}{\max}\left(\alpha_{T}\left(v\right)\right)$
- obtained by iterating the algorithm on the hb-graph - in incremental
steps and while the eccentricity is kept above 0. In order to have
a ratio we calculate: 
\[
r_{V}=\dfrac{s_{V}}{\alpha_{\text{ref}}}
\]
 where $\alpha_{\text{ref}}$ is the reference normalised value used
for the initialisation of the $\alpha$ value of the vertices of the
hb-graph $\mathcal{H}$. This ratio has values increasing by steps
from 0 to $\dfrac{\alpha_{T,\text{max}}}{\alpha_{\text{ref}}}$.

\begin{figure}
\begin{center}\includegraphics[scale=0.35]{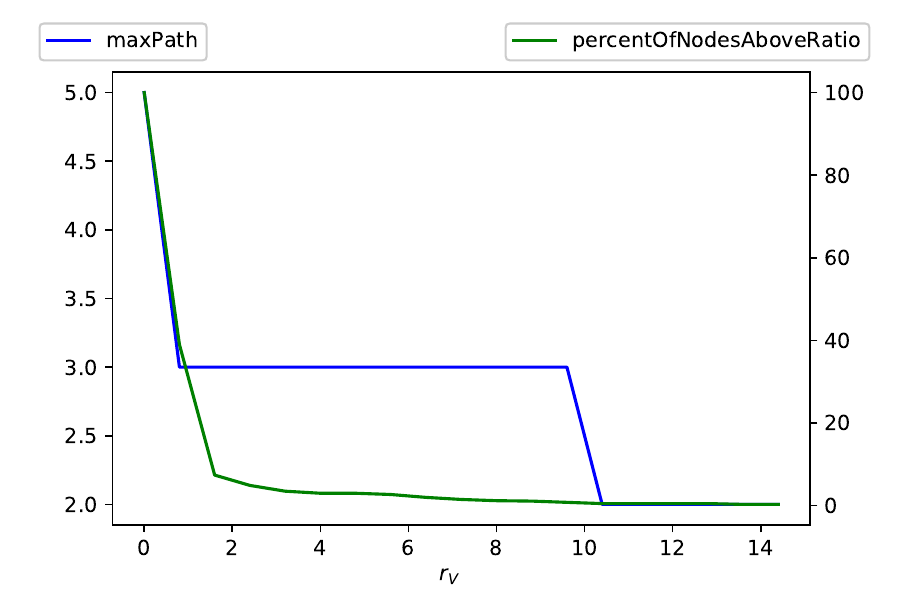}\end{center}

\caption{Maximum path length and percentage of vertices in $A_{V}(s)$ over
vertices in $V$ vs ratio $r_{V}.$}
\label{Fig:vertices_max_path}
\end{figure}

We show the results obtained in Figure \ref{Fig:vertices_max_path}
on two plots. The first plot corresponds to the maximal length of
the path between vertices of $A_{V}\left(s_{V}\right)$ and vertices
of $B_{V}\left(s_{V}\right)$ that are connected according to the
ratio $r_{V}=\dfrac{s_{V}}{\alpha_{\text{ref}}}$: this path length
corresponds to half of the length of the path observed in the extra-vertex
graph representation of the hb-graph support hypergraph as in between
two vertices of $V$ there is an extra-vertex that represents the
hb-edge (or the support hyperedge). The second curve plots the percentage
of vertices of $V$ that are in $A_{V}\left(s_{V}\right)$ in function
of $r_{V}$. When $r_{V}$ increases the number of elements in $A_{V}\left(s_{V}\right)$
naturally decreases while they get closer to the elements of $B_{V}\left(s_{V}\right)$,
marking the fact that they are central.

\begin{figure}
\begin{center}\includegraphics[scale=0.42]{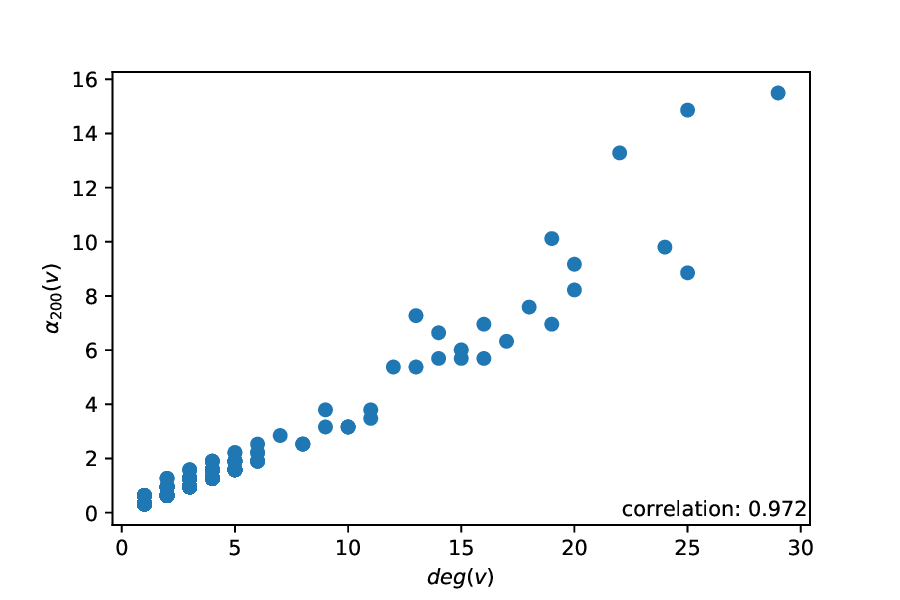}\end{center}

\caption{Alpha value of vertices at step 200 and degree of vertices.}
\label{Fig:vertices_alpha_val_degree}
\end{figure}

\begin{figure}
\begin{center}\includegraphics[scale=0.42]{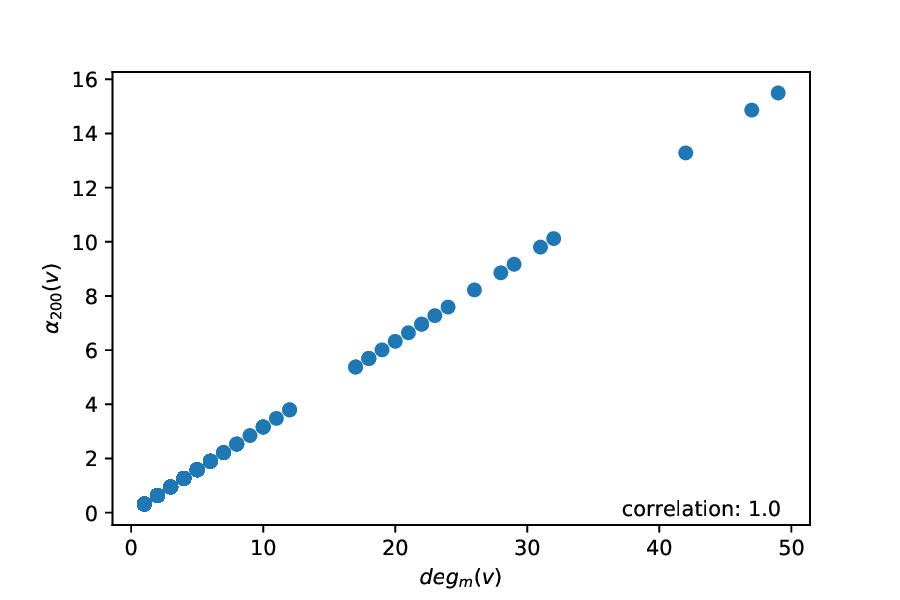}\end{center}

\caption{Alpha value of vertices at step 200 and m-degree of vertices.}
\label{Fig:vertices_alpha_val_mdegree}
\end{figure}

\begin{figure*}
\begin{center}\includegraphics[scale=0.35]{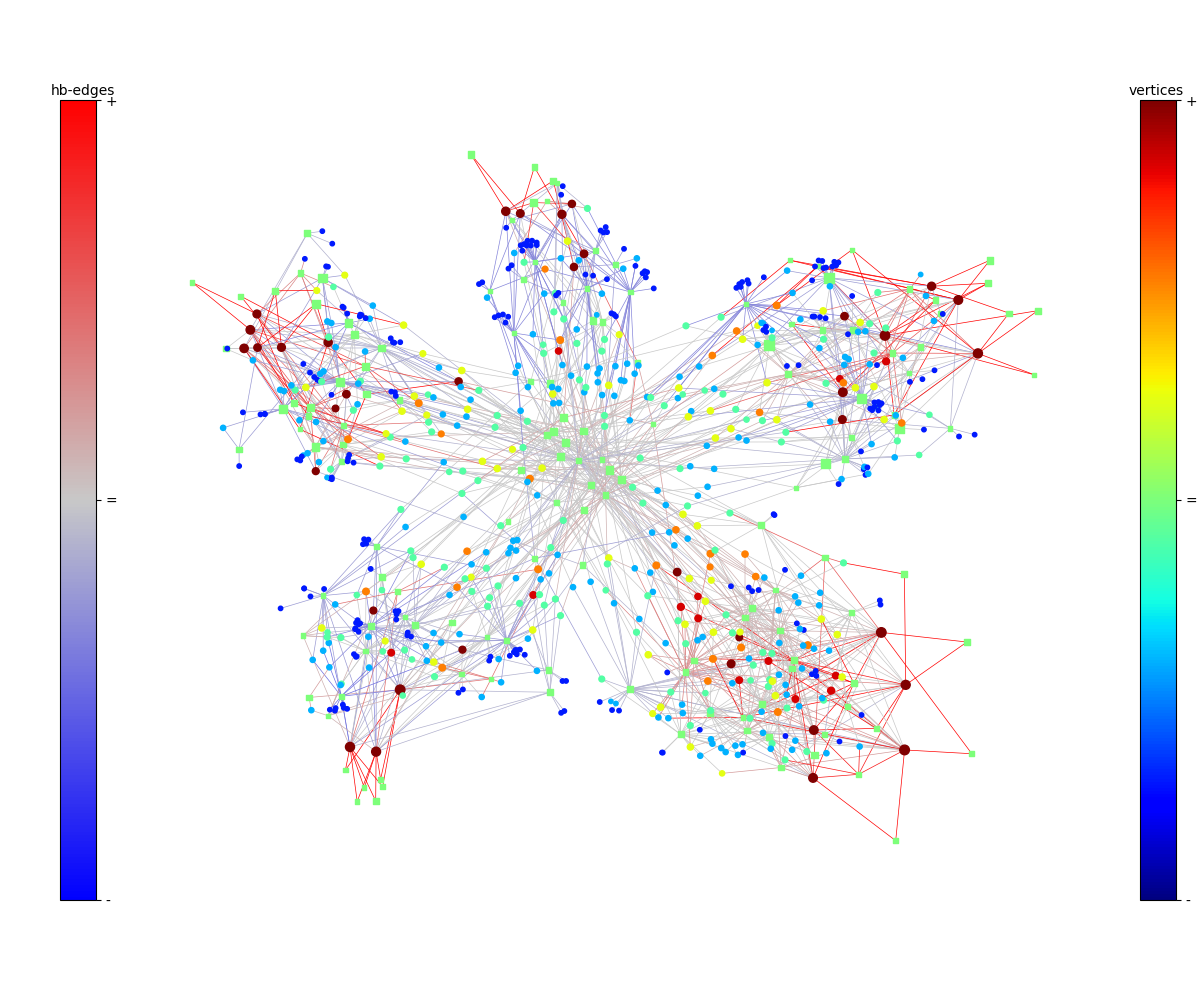}\end{center}

\caption{Exchange-based diffusion in hb-graphs after 200 iterations of Algorithm
\ref{Alg: Exchange}: highlighting important hb-edges. Simulation
with 807 vertices (chosen randomly out of 10~000) gathered in 5 groups
of vertices (with 6, 5, 7, 3 and 5 important vertices and 2 important
vertices per hb-edge), 220 hb-edges (with cardinality of support less
or equal to 25), 20 vertices in between the 5 groups. Extra-vertices
are colored in green and have square shape.}
\label{Fig:Exchanges}
\end{figure*}

Figure \ref{Fig:vertices_alpha_val_degree} and Figure \ref{Fig:vertices_alpha_val_mdegree}
show that high values of $\alpha_{T}\left(v\right)$ correspond to
vertices that are highly connected either by degree or by m-degree.
Hence vertices in Figure \ref{Fig:Exchanges} that are on the positive
side of the scale color correspond to highly connected vertices: the
closer to red on the right scale they are, the higher the value of
$\alpha_{T}\left(v\right)$ is.

A similar approach is taken for the hb-edges: assuming that the diffusion
process stops at time $T$, we use the $\epsilon_{T-\frac{1}{2}}$
function to partition the set of hb-edges into two subsets for a given
threshold $s_{E}$: $A_{E}\left(s_{E}\right)$ of the hb-edges that
have $\epsilon$ values above the threshold and $B_{E}\left(s_{E}\right)$
the one gathering hb-edges that have $\epsilon$ values below $s_{E}$.

\begin{figure*}
\begin{center}\includegraphics[scale=0.42]{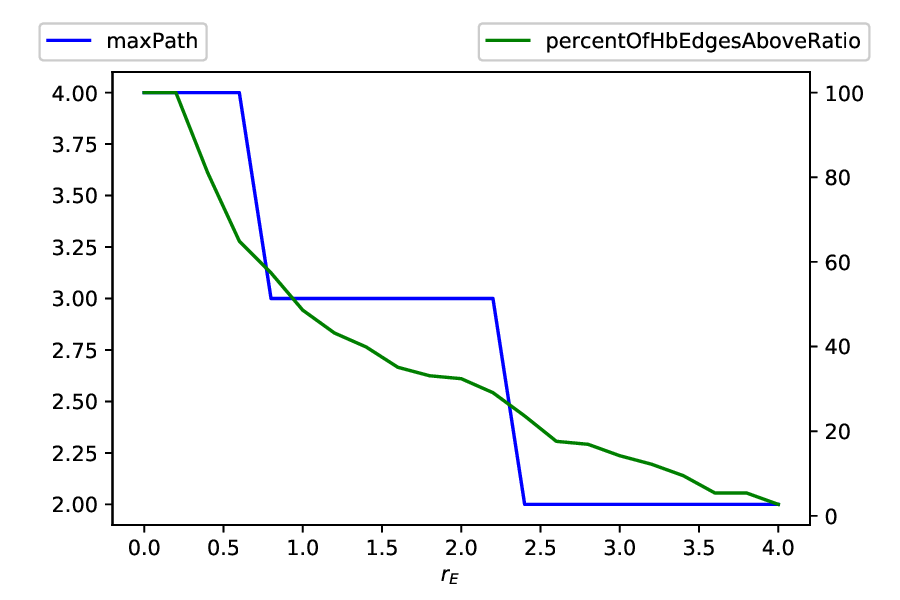}\end{center}

\caption{Path maximum length and percentage of vertices in $A_{E}(s)$ vs ratio.}
\label{Fig:hb_edge_max_path}
\end{figure*}

$s_{E}$ varies from 0 to $\epsilon_{T-\frac{1}{2},\text{max}}=\underset{e\in E}{\max}\left(\epsilon_{T-\frac{1}{2}}\left(e\right)\right)$
by incremental steps while keeping the eccentricity above 0, first
of the two conditions achieved. In the hb-graph representation, each
hb-edge corresponds to an extra-vertex. Each time we evaluate the
length of the maximal shortest path linking one vertex of $A_{E}\left(s_{E}\right)$
to one vertex of $B_{E}\left(s_{E}\right)$ for connected vertices
in the hb-graph support hypergraph extra-vertex graph representation:
the length of the path corresponds to half of the one obtained from
the graph for the same reason than before. \textcolor{blue}{We define
the ratio 
\[
r_{E}=\dfrac{s_{E}}{\beta_{\text{ref}}}
\]
 where $\beta_{\text{ref}}=\dfrac{1}{\left|E\right|}$ that corresponds
to the normalised value that would be used in the dual hb-graph to
initialise the diffusion process.} In Figure \ref{Fig:hb_edge_max_path},
we observe for the hb-edges the same trend than the one observed for
vertices: the length of the maximal path between two hb-edges decreases
as the ratio $r_{E}$ increases while the percentage of vertices in
$A_{E}\left(s_{E}\right)$ over $V$ decreases.

\begin{figure}
\begin{center}\includegraphics[scale=0.42]{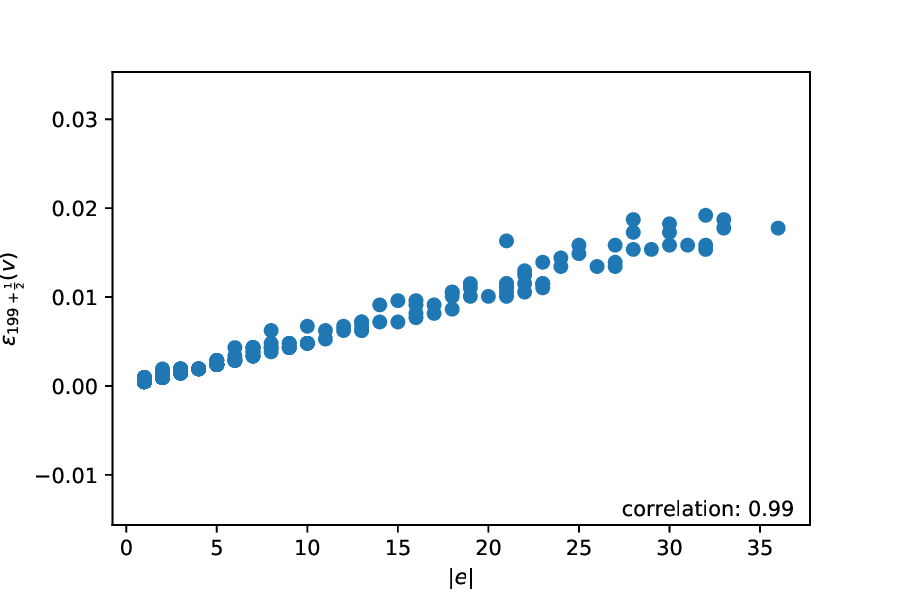}\end{center}

\caption{Epsilon value of hb-edge at stage 199+$\frac{1}{2}$ and cardinality
of hb-edge.}
\label{Fig:hb-edge_cardinality}
\end{figure}

\begin{figure}
\begin{center}\includegraphics[scale=0.42]{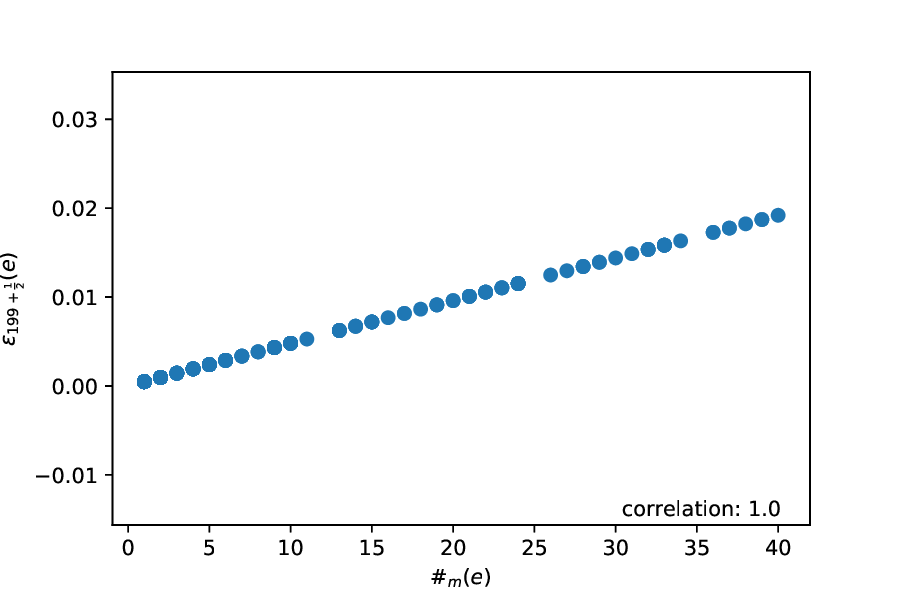}\end{center}

\caption{Epsilon value of hb-edge at stage 199+$\frac{1}{2}$ and (m-)cardinality
of hb-edge.}
\label{Fig:hb-edge_mcardinality}
\end{figure}

Figure \ref{Fig:hb-edge_cardinality} shows the high correlation between
the value of $\epsilon(e)$ and the cardinality of $e$; Figure \ref{Fig:hb-edge_mcardinality}
shows that the correlation between value of $\epsilon(e)$ and the
m-cardinality of $e$ is even stronger.

\begin{figure}
\begin{center}\includegraphics[scale=0.35]{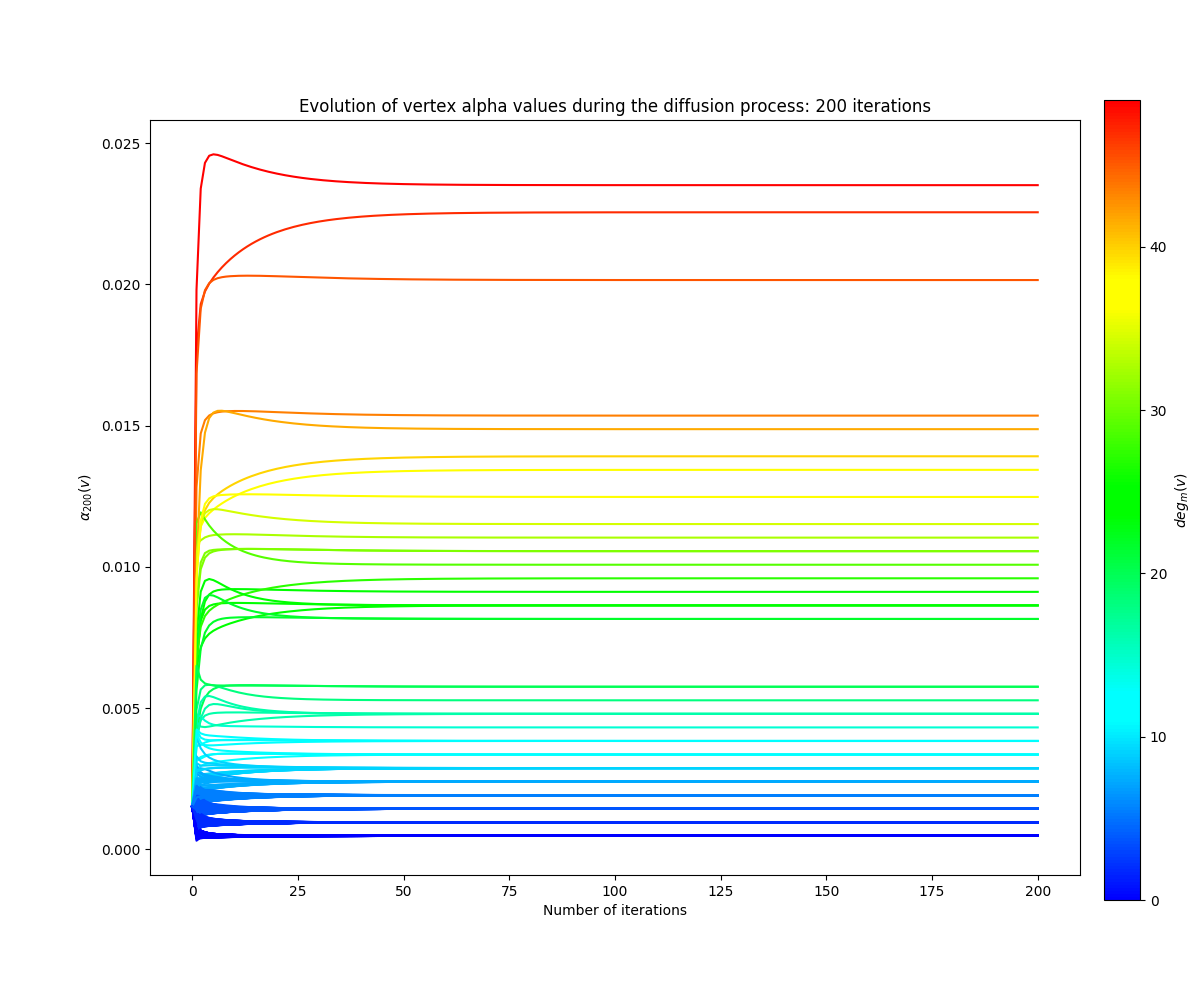}\end{center}

\caption{Alpha value convergence of the vertices vs number of iterations. The
plots are m-degree-based gradiently colored.}
\label{Fig:vertex alpha value cv}
\end{figure}

\begin{figure}
\begin{center}\includegraphics[scale=0.35]{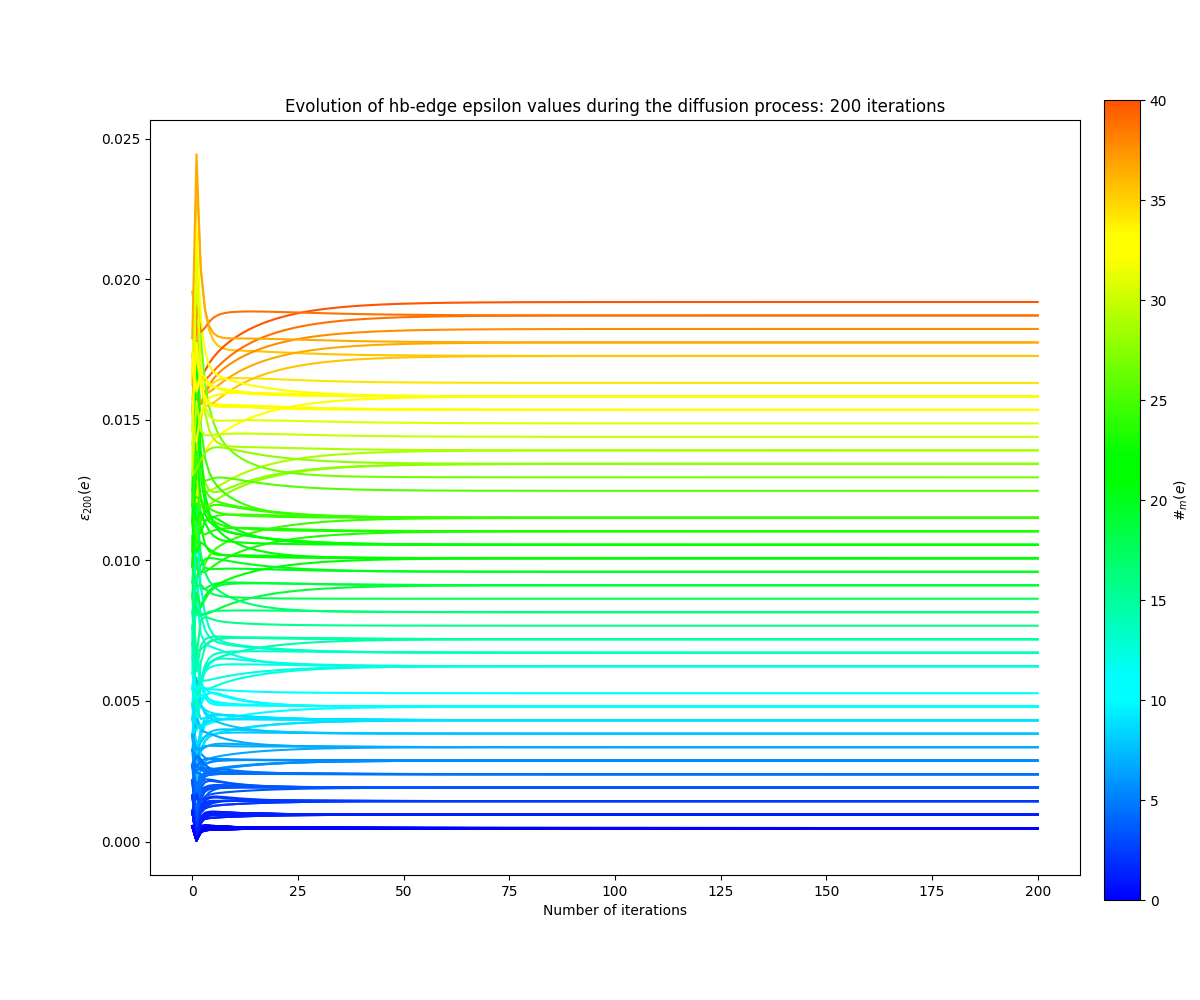}\end{center}

\caption{Epsilon value convergence of hb-edges vs number of iterations. The
plots are m-cardinality-based gradiently colored.}
\label{Fig:hb-edge_epsilon_value_cv}
\end{figure}

\textcolor{blue}{The number of iterations needed to have a significant
convergence depends on the initial conditions; we tried different
initialisations, either uniform, or applying some strokes on a different
number of nodes. We observed that the more uniform the information
on the network is, the less number of iterations for convergence is
needed. No matter the configuration, the most important vertices in
term of connectivity are always the most highlighted. Figure \ref{Fig:vertex alpha value cv}
and in Figure \ref{Fig:hb-edge_epsilon_value_cv} depict the convergence
observed on a uniform initial distribution as it is described in the
former section. In Figure \ref{Fig:vertex alpha value cv}, we can
see how the $\alpha$-values as we observed in Figure \ref{Fig:vertices_alpha_val_degree}
reflect the m-degree of the vertex they are associated to: 200 iterations
is far enough to rank the vertices by m-degree. In Figure \ref{Fig:hb-edge_epsilon_value_cv}
we can observe an analoguous phenomena with the $\epsilon$-value
associated to hb-edges that reflect the m-cardinality of the hb-edges.
Again 200 iterations are sufficient to converge in studied cases.}

\textcolor{blue}{The iterations needed to converge depends on the
structure of the network. In the transitory phase, the vertices need
to exchange with the hb-edges; the process requires some iterations
before converging and its behaviour depends on the node connectivity
and the hb-edge composition. It is an open question to investigate
on this transitory phase to have more indications on the way the $\epsilon$
and the $\alpha$-values vary.}

As we already mentioned the results on hb-edges show that the values
obtained are highly correlated to the m-cardinality of the hyperedges.
To color the hb-edges as it is done in Figure \ref{Fig:Exchanges},
we calculate the ratio $r_{T-\frac{1}{2}}\left(e\right)=\dfrac{\epsilon_{T-\frac{1}{2}}\left(e\right)}{\epsilon_{\text{norm}}(e)}$,
where $\epsilon_{\text{norm}}\left(e\right)=\sum\limits _{v\in e^{\star}}\dfrac{m_{e}\left(v\right)}{\deg_{m}(v)}v_{\text{ref}}$
corresponds to the value obtained from the vertices of the hb-edge
support by giving to each of them the reference value. Hb-edges are
colored using $r_{T-\frac{1}{2}}\left(e\right)$, the higher the value,
the closer to red the color of the left gradient color bar is.

\textcolor{blue}{To compare our exchange-based diffusion process to
a baseline we considered a classical random walk. In this classical
random walk, the walker who is on a vertex $v$ chooses randomly a
hb-edge that is incident with a uniform probability law and when the
walker is on a hb-edge $e$ he chooses a vertex inside the hb-edge
randomly with a uniform probability law. We let the possibility of
teleportation to an other vertex from a vertex with a tunable value
$\gamma$: $1-\gamma$ represents the probability to be teleported.
We choose $\gamma=0.85$. We count the number of passages of the walker
through each vertex and each hb-edge. We stop the random walk when
the hb-graph is fully explored. We iterate $N$ times the random walk,
$N$ varying.}

\textcolor{blue}{To improve the results of the classical random walk
we propose a modified random walk - described in Algorithm \ref{Alg: RW}
-} on the hb-graphs with random choice of hb-edges when the walker
is on a vertex $v$ with a distribution of probability $\left(\dfrac{w_{e}\left(e_{i}\right)m_{i}\left(v\right)}{\deg_{w,m}\left(v\right)}\right)_{1\leqslant i\leqslant p}$
and a random choice of the vertex when the walker is on a hb-edge
$e$ with a distribution of probability $\left(\dfrac{m_{e}\left(v_{i}\right)}{\#_{m}\left(e\right)}\right)_{1\leqslant i\leqslant n}.$
We let the possibility of teleportation as it is done in the classical
random walk. Similarly to the classical random walk, we count the
number of passages of the walker through each vertex and each hb-edge.
We also stop the random walk when the hb-graph is fully explored.
We iterate $N$ times the random walk with various values of $N.$
Assigning a multiplicity of 1 to every vertex and a weight of 1 for
every hb-edge - with the vertex degree and the hb-edge cardinality
instead of the multiplicity - retrieves the classical random walk
from the modified random walk.

\begin{algorithm}
\textbf{\textcolor{blue}{Given:}}

\textcolor{blue}{\hspace{0.5cm}A hb-graph $\mathcal{H}=\left(V,E,w_{e}\right)$
with $\left|V\right|=n$ and $\left|E\right|=p$}

\textcolor{blue}{\hspace{0.5cm}Number of Random walks: $T_{\text{RW}}$}

\textcolor{blue}{\hspace{0.5cm}A teleportation threshold: $\gamma_{\text{th}}$}\\

\textbf{\textcolor{blue}{Initialisation:}}

\textcolor{blue}{\hspace{0.5cm}$\forall v\in V:$ $n_{V}\left(v\right)=0$}

\textcolor{blue}{\hspace{0.5cm}$\forall e\in E$~:$n_{E}\left(e\right)=0$}

\textcolor{blue}{\hspace{0.5cm}Q := deep copy $\left(V\right)$}

\textcolor{blue}{\hspace{0.5cm}$v_{0}$ := random $(v\in Q)$}

\textcolor{blue}{\hspace{0.5cm}$n_{V}\left(v_{0}\right)=1$}

\textcolor{blue}{\hspace{0.5cm}$Q:=Q\backslash\left\{ v_{0}\right\} $}\\

\textbf{\textcolor{blue}{OneRW():}}

\textcolor{blue}{\hspace{0.5cm}While $Q\neq\emptyset$:}

\textcolor{blue}{\hspace{0.5cm}\hspace{0.5cm}$\gamma_{\text{rand}}=\text{random}\left(\left[0;1\right],\text{weight}=\text{uniform}\right)$}

\textcolor{blue}{\hspace{0.5cm}\hspace{0.5cm}if $\gamma_{\text{rand}}<\gamma_{\text{th}}$:}

\textcolor{blue}{\hspace{0.5cm}\hspace{0.5cm}\hspace{0.5cm}\# Visit
of incident edges}

\textcolor{blue}{\hspace{0.5cm}\hspace{0.5cm}\hspace{0.5cm}$e_{\text{c}}:=\text{random}\left(e\in E:v_{\text{c}}\in e^{\star},\text{weight}=\left(\dfrac{w_{e}\left(e_{j}\right)m_{e_{j}}\left(v_{0}\right)}{\deg_{w_{e},m}\left(v_{0}\right)}\right)_{e_{j}\in E}\right)$}

\textcolor{blue}{\hspace{0.5cm}\hspace{0.5cm}\hspace{0.5cm}$n_{V}\left(e_{\text{c}}\right):=n_{V}\left(e_{\text{c}}\right)+1$}

\textcolor{blue}{\hspace{0.5cm}\hspace{0.5cm}\hspace{0.5cm}\# Choice
of the next vertex}

\textcolor{blue}{\hspace{0.5cm}\hspace{0.5cm}\hspace{0.5cm}$v_{0}:=\text{random}\left(v\in V:v\in e_{\text{c}}^{\star},\text{weight}=\left(\dfrac{m_{e_{\text{c}}}\left(v\right)}{\#_{m}\left(e_{\text{c}}\right)}\right)_{v\in V}\right)$}

\textcolor{blue}{\hspace{0.5cm}\hspace{0.5cm}\hspace{0.5cm}If $v_{0}\in Q$:}

\textcolor{blue}{\hspace{0.5cm}\hspace{0.5cm}\hspace{0.5cm}\hspace{0.5cm}$Q:=Q\backslash\left\{ v_{0}\right\} $}

\textcolor{blue}{\hspace{0.5cm}\hspace{0.5cm}\hspace{0.5cm}$n_{V}\left(v_{0}\right):=n_{V}\left(v_{0}\right)+1$}

\textcolor{blue}{\hspace{0.5cm}\hspace{0.5cm}else:}

\textcolor{blue}{\hspace{0.5cm}\hspace{0.5cm}\hspace{0.5cm}\# Case
of teleportation}

\textcolor{blue}{\hspace{0.5cm}\hspace{0.5cm}\hspace{0.5cm}$v_{0}:=\text{random}\left(v\in V:v\in e_{\text{c}}^{\star},\text{weight}=\left(\dfrac{m_{e_{\text{c}}}\left(v\right)}{\#_{m}\left(e_{\text{c}}\right)}\right)_{v\in V}\right)$}

\textcolor{blue}{\hspace{0.5cm}\hspace{0.5cm}\hspace{0.5cm}$Q:=Q\backslash\left\{ v_{0}\right\} $}

\textcolor{blue}{\hspace{0.5cm}\hspace{0.5cm}\hspace{0.5cm}$n_{V}\left(v_{0}\right):=n_{V}\left(v_{0}\right)+1$}\\
\textcolor{blue}{}\\
\textbf{\textcolor{blue}{Main():}}

\textcolor{blue}{\hspace{0.5cm}For i:=0 to $T_{\text{RW}}:$}

\textcolor{blue}{\hspace{0.5cm}\hspace{0.5cm}OneRW()}

\textcolor{blue}{\hspace{0.5cm}$\forall v\in V$: $\overline{n_{V}}\left(v\right)=\dfrac{n_{V}\left(v\right)}{T_{\text{RW}}}$}

\textcolor{blue}{\hspace{0.5cm}$\forall e\in E$: $\overline{n_{E}}\left(e\right)=\dfrac{n_{E}\left(e\right)}{T_{\text{RW}}}$}

\caption{Modified random walk in hb-graphs}

\label{Alg: RW}
\end{algorithm}

\begin{figure}
\begin{center}\includegraphics[scale=0.42]{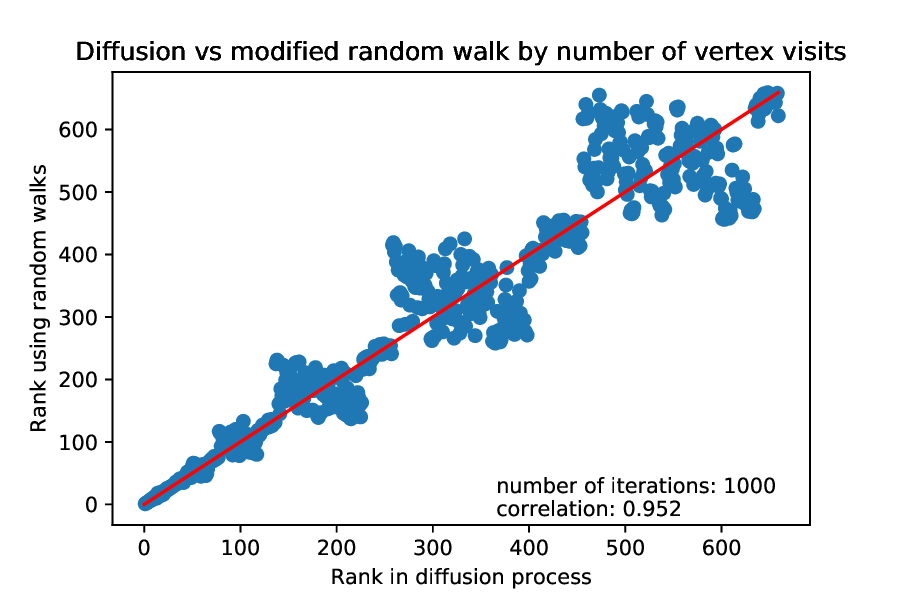}\end{center}

\caption{Comparison of the rank obtained by a thousand modified random walks
after total discovery of the vertices in the hb-graph and rank obtained
with 200 iterations of the exchange-based diffusion process.}
\label{Fig:ModifiedRW_rank_vs_diff_rank}
\end{figure}

\begin{figure}
\begin{center}\includegraphics[scale=0.42]{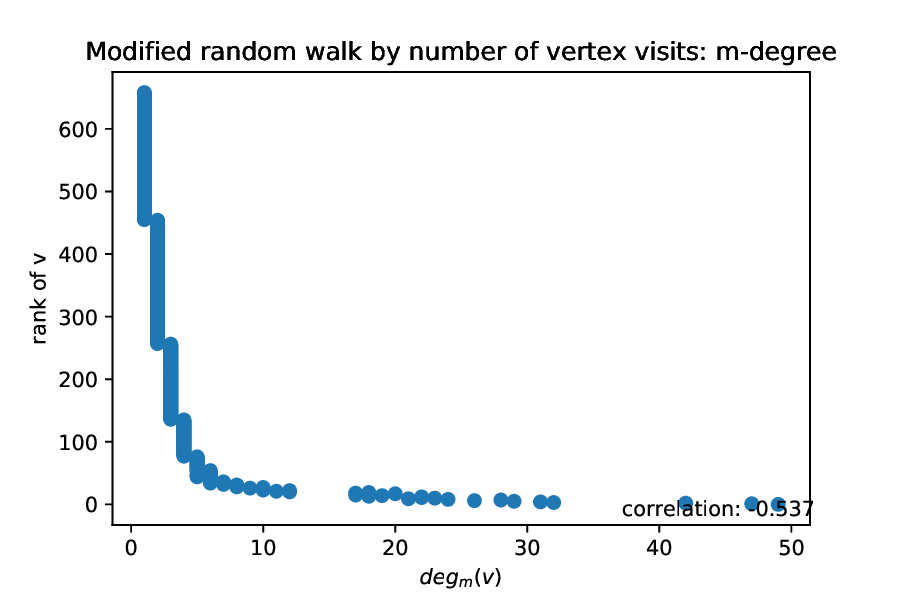}\end{center}

\caption{Comparison of the rank obtained by a thousand modified random walks
after total discovery of the vertices in the hb-graph and m-degree
of vertices}
\label{Fig:ModifiedRW_rank_vs_deg}
\end{figure}

\begin{figure}
\begin{center}\includegraphics[scale=0.42]{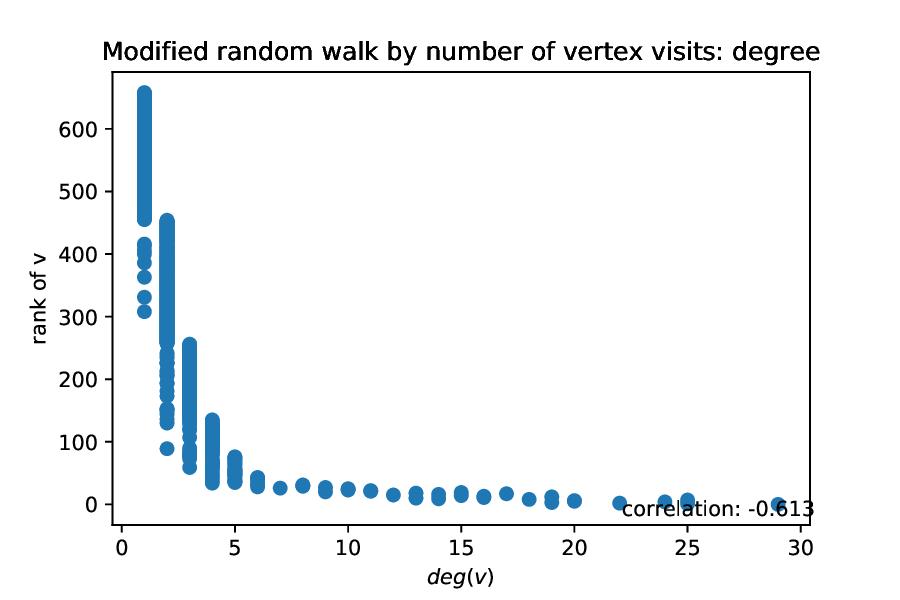}\end{center}

\caption{Comparison of the rank obtained by a thousand modified random walks
after total discovery of the vertices in the hb-graph and degree of
vertices}
\label{Fig:ModifiedRW_rank_vs_mdeg}
\end{figure}

\textcolor{blue}{Figure \ref{Fig:ModifiedRW_rank_vs_diff_rank} shows
that there is a good correlation between the rank obtained by a thousand
modified random walks and after two hundreds iterations of our diffusion
process, especially for the first hundred vertices of the network,
which is generally the ones that are targetted. The lack of correlation
between the rank obtained by the random walk with the degree of the
vertices and the m-degree of vertices as shown respectively in Figure
\ref{Fig:ModifiedRW_rank_vs_deg} and Figure \ref{Fig:ModifiedRW_rank_vs_mdeg}
is mainly due to the vertices with low m-degrees / degrees.}

\begin{figure}
\begin{center}\includegraphics[scale=0.42]{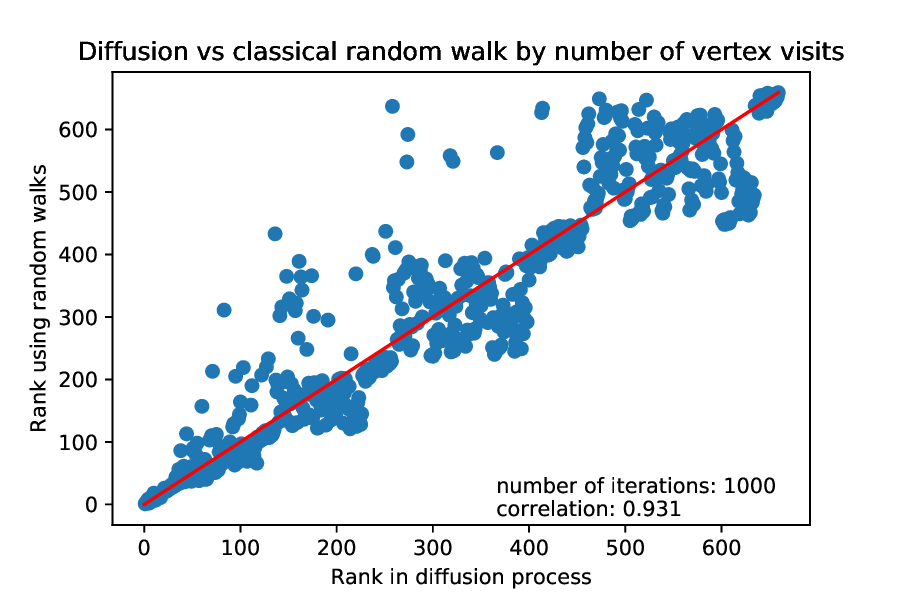}\end{center}

\caption{Comparison of the rank obtained by a thousand classical random walks
after total discovery of the vertices in the hb-graph and rank obtained
with 200 iterations of the exchange-based diffusion process.}
\label{Fig:ClassicalRW_rank_vs_diff_rank}
\end{figure}

\begin{figure}
\begin{center}\includegraphics[scale=0.3]{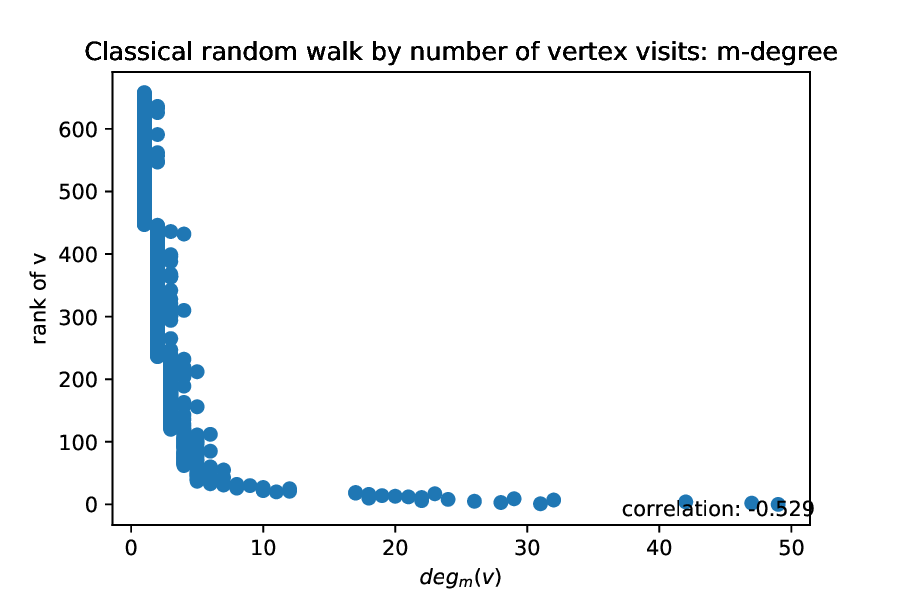}\end{center}

\caption{Comparison of the rank obtained by a thousand classical random walks
after total discovery of the vertices in the hb-graph and m-degree
of vertices}
\label{Fig:ClassicalRW_rank_vs_deg}
\end{figure}

\begin{figure}
\begin{center}\includegraphics[scale=0.3]{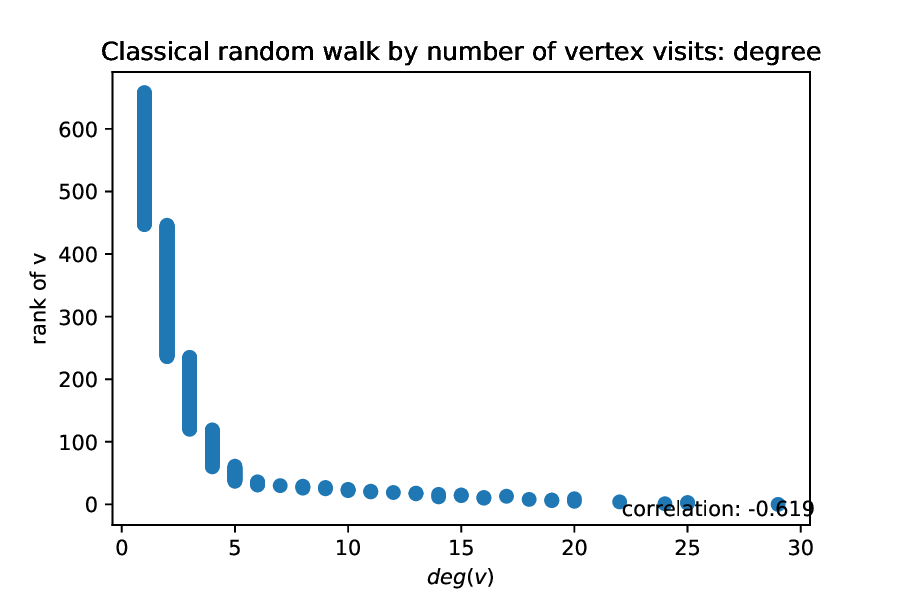}\end{center}

\caption{Comparison of the rank obtained by a thousand classical random walks
after total discovery of the vertices in the hb-graph and degree of
vertices}
\label{Fig:ClassicalRW_rank_vs_mdeg}
\end{figure}

\textcolor{blue}{We can remark in Figure \ref{Fig:ClassicalRW_rank_vs_diff_rank}
that the correlation is a bit lower with a thousand classical random
walks due to the fact that there are more vertices that are seen as
differently ranked in between the two approaches. In Figure \ref{Fig:ClassicalRW_rank_vs_deg},
we can see that the ranks in the classical random walk relies more
on the degree than on the m-degree as shown in Figure \ref{Fig:ClassicalRW_rank_vs_mdeg},
especially for vertices with small (m-)degrees; but there is still
a misclassification for lower (m-)degree vertices.}

\textcolor{blue}{We have compared the three methods from a computational
time perspective; the results are shown in Table \ref{Tab: Comparison of time}.
The diffusion process is clearly faster; the modified random walk,
essentially due to the overhead due to the large number of divisions,
requires longer than the classical random walk. A lot of optimisation
can be foreseen to make this modified random walk running faster.
The random walks can be easily parallelised; it is also the case for
the diffusion process. The number of iterations in the diffusion process
can also be optimised. These issues will be addressed in future work.}

\begin{table}
\begin{center}\scalebox{0.75}{\textcolor{blue}{}%
\begin{tabular}{|c|c|c|c|c|c|c|c|c|c|}
\hline 
\textcolor{blue}{$\left|E\right|$} & \textcolor{blue}{$\left|V\right|$} & \textcolor{blue}{$k$} & \textcolor{blue}{$N_{1}$} & \textcolor{blue}{$N_{0}$} & \textcolor{blue}{Type of algorithm} & \textcolor{blue}{100} & \textcolor{blue}{200} & \textcolor{blue}{500} & \textcolor{blue}{1000}\tabularnewline
\hline 
\textcolor{blue}{55} & \textcolor{blue}{106} & \textcolor{blue}{1} & \textcolor{blue}{5} & \textcolor{blue}{5} & \textcolor{blue}{classical random walk} & \textcolor{blue}{0.40 \textpm{} 0.05} & \textcolor{blue}{0.78 \textpm{} 0.07} & \textcolor{blue}{1.92 \textpm{} 0.10} & \textcolor{blue}{3.82 \textpm{} 0.14}\tabularnewline
\hline 
\textcolor{blue}{55} & \textcolor{blue}{106} & \textcolor{blue}{1} & \textcolor{blue}{5} & \textcolor{blue}{5} & \textcolor{blue}{diffusion} & \textcolor{blue}{0.05 \textpm{} 0.02} & \textcolor{blue}{0.08 \textpm{} 0.02} & \textcolor{blue}{0.20 \textpm{} 0.04} & \textcolor{blue}{0.39 \textpm{} 0.06}\tabularnewline
\hline 
\textcolor{blue}{55} & \textcolor{blue}{106} & \textcolor{blue}{1} & \textcolor{blue}{5} & \textcolor{blue}{5} & \textcolor{blue}{modified random walk} & \textcolor{blue}{0.71 \textpm{} 0.06} & \textcolor{blue}{1.43 \textpm{} 0.09} & \textcolor{blue}{3.56 \textpm{} 0.17} & \textcolor{blue}{7.12 \textpm{} 0.23}\tabularnewline
\hline 
\textcolor{blue}{55} & \textcolor{blue}{132} & \textcolor{blue}{3} & \textcolor{blue}{5} & \textcolor{blue}{5} & \textcolor{blue}{classical random walk} & \textcolor{blue}{0.49 \textpm{} 0.05} & \textcolor{blue}{0.96 \textpm{} 0.06} & \textcolor{blue}{2.36 \textpm{} 0.08} & \textcolor{blue}{4.71 \textpm{} 0.12}\tabularnewline
\hline 
\textcolor{blue}{55} & \textcolor{blue}{132} & \textcolor{blue}{3} & \textcolor{blue}{5} & \textcolor{blue}{5} & \textcolor{blue}{diffusion} & \textcolor{blue}{0.05 \textpm{} 0.02} & \textcolor{blue}{0.09 \textpm{} 0.02} & \textcolor{blue}{0.21 \textpm{} 0.04} & \textcolor{blue}{0.42 \textpm{} 0.05}\tabularnewline
\hline 
\textcolor{blue}{55} & \textcolor{blue}{132} & \textcolor{blue}{3} & \textcolor{blue}{5} & \textcolor{blue}{5} & \textcolor{blue}{modified random walk} & \textcolor{blue}{0.89 \textpm{} 0.06} & \textcolor{blue}{1.77 \textpm{} 0.09} & \textcolor{blue}{4.43 \textpm{} 0.13} & \textcolor{blue}{8.85 \textpm{} 0.19}\tabularnewline
\hline 
\textcolor{blue}{55} & \textcolor{blue}{91} & \textcolor{blue}{5} & \textcolor{blue}{5} & \textcolor{blue}{5} & \textcolor{blue}{classical random walk} & \textcolor{blue}{0.30 \textpm{} 0.04} & \textcolor{blue}{0.59 \textpm{} 0.05} & \textcolor{blue}{1.44 \textpm{} 0.06} & \textcolor{blue}{2.85 \textpm{} 0.07}\tabularnewline
\hline 
\textcolor{blue}{55} & \textcolor{blue}{91} & \textcolor{blue}{5} & \textcolor{blue}{5} & \textcolor{blue}{5} & \textcolor{blue}{diffusion} & \textcolor{blue}{0.04 \textpm{} 0.02} & \textcolor{blue}{0.07 \textpm{} 0.02} & \textcolor{blue}{0.16 \textpm{} 0.03} & \textcolor{blue}{0.31 \textpm{} 0.04}\tabularnewline
\hline 
\textcolor{blue}{55} & \textcolor{blue}{91} & \textcolor{blue}{5} & \textcolor{blue}{5} & \textcolor{blue}{5} & \textcolor{blue}{modified random walk} & \textcolor{blue}{0.55 \textpm{} 0.05} & \textcolor{blue}{1.09 \textpm{} 0.06} & \textcolor{blue}{2.71 \textpm{} 0.09} & \textcolor{blue}{5.42 \textpm{} 0.14}\tabularnewline
\hline 
\textcolor{blue}{305} & \textcolor{blue}{534} & \textcolor{blue}{1} & \textcolor{blue}{5} & \textcolor{blue}{5} & \textcolor{blue}{classical random walk} & \textcolor{blue}{4.05 \textpm{} 0.16} & \textcolor{blue}{8.07 \textpm{} 0.26} & \textcolor{blue}{20.10 \textpm{} 0.45} & \textcolor{blue}{40.17 \textpm{} 0.85}\tabularnewline
\hline 
\textcolor{blue}{305} & \textcolor{blue}{534} & \textcolor{blue}{1} & \textcolor{blue}{5} & \textcolor{blue}{5} & \textcolor{blue}{diffusion} & \textcolor{blue}{0.29 \textpm{} 0.06} & \textcolor{blue}{0.57 \textpm{} 0.08} & \textcolor{blue}{1.35 \textpm{} 0.09} & \textcolor{blue}{2.64 \textpm{} 0.10}\tabularnewline
\hline 
\textcolor{blue}{305} & \textcolor{blue}{534} & \textcolor{blue}{1} & \textcolor{blue}{5} & \textcolor{blue}{5} & \textcolor{blue}{modified random walk} & \textcolor{blue}{6.86 \textpm{} 0.28} & \textcolor{blue}{13.71 \textpm{} 0.41} & \textcolor{blue}{34.16 \textpm{} 0.75} & \textcolor{blue}{68.28 \textpm{} 1.21}\tabularnewline
\hline 
\textcolor{blue}{305} & \textcolor{blue}{491} & \textcolor{blue}{3} & \textcolor{blue}{5} & \textcolor{blue}{5} & \textcolor{blue}{classical random walk} & \textcolor{blue}{3.51 \textpm{} 0.13} & \textcolor{blue}{6.98 \textpm{} 0.21} & \textcolor{blue}{17.39 \textpm{} 0.38} & \textcolor{blue}{34.77 \textpm{} 0.70}\tabularnewline
\hline 
\textcolor{blue}{305} & \textcolor{blue}{491} & \textcolor{blue}{3} & \textcolor{blue}{5} & \textcolor{blue}{5} & \textcolor{blue}{diffusion} & \textcolor{blue}{0.27 \textpm{} 0.05} & \textcolor{blue}{0.53 \textpm{} 0.09} & \textcolor{blue}{1.25 \textpm{} 0.11} & \textcolor{blue}{2.43 \textpm{} 0.11}\tabularnewline
\hline 
\textcolor{blue}{305} & \textcolor{blue}{491} & \textcolor{blue}{3} & \textcolor{blue}{5} & \textcolor{blue}{5} & \textcolor{blue}{modified random walk} & \textcolor{blue}{6.02 \textpm{} 0.22} & \textcolor{blue}{12.03 \textpm{} 0.41} & \textcolor{blue}{30.10 \textpm{} 0.73} & \textcolor{blue}{60.23 \textpm{} 1.34}\tabularnewline
\hline 
\textcolor{blue}{305} & \textcolor{blue}{499} & \textcolor{blue}{5} & \textcolor{blue}{5} & \textcolor{blue}{5} & \textcolor{blue}{classical random walk} & \textcolor{blue}{3.31 \textpm{} 0.15} & \textcolor{blue}{6.58 \textpm{} 0.20} & \textcolor{blue}{16.38 \textpm{} 0.34} & \textcolor{blue}{32.72 \textpm{} 0.51}\tabularnewline
\hline 
\textcolor{blue}{305} & \textcolor{blue}{499} & \textcolor{blue}{5} & \textcolor{blue}{5} & \textcolor{blue}{5} & \textcolor{blue}{diffusion} & \textcolor{blue}{0.24 \textpm{} 0.04} & \textcolor{blue}{0.47 \textpm{} 0.06} & \textcolor{blue}{1.12 \textpm{} 0.06} & \textcolor{blue}{2.18 \textpm{} 0.08}\tabularnewline
\hline 
\textcolor{blue}{305} & \textcolor{blue}{499} & \textcolor{blue}{5} & \textcolor{blue}{5} & \textcolor{blue}{5} & \textcolor{blue}{modified random walk} & \textcolor{blue}{5.86 \textpm{} 0.26} & \textcolor{blue}{11.70 \textpm{} 0.37} & \textcolor{blue}{29.26 \textpm{} 0.58} & \textcolor{blue}{58.51 \textpm{} 0.89}\tabularnewline
\hline 
\end{tabular}}\end{center}

\caption{\textcolor{blue}{Time taken for doing 100, 200, 500 and 1000 iterations
of the diffusion algorithm and 100, 200, 500 and 1000 classical and
modified random walks on different hb-graphs}}

\label{Tab: Comparison of time}
\end{table}

\subsection{\textcolor{blue}{Application to Arxiv querying}}

\textcolor{blue}{We used the standard Arxiv API}\footnote{\textcolor{blue}{\href{https://arxiv.org/help/api/index}{https://arxiv.org/help/api/index}}}\textcolor{blue}{{}
to perform searches on Arxiv database. When performing a search, the
query is transformed into a vector of words which is the basis for
the retrieval of documents. The most relevant documents are retrieved
based on a similarity measure between the query vector and the word
vectors associated to individual documents. Arxiv relies on Lucene's
built-in Vector Space Model of information retrieval and the boolean
model}\footnote{\textcolor{blue}{\href{https://lucene.apache.org/core/2_9_4/scoring.html}{https://lucene.apache.org/core/2\_9\_4/scoring.html}}}\textcolor{blue}{.
The Arxiv API returns the metadata associated to the document with
highest scores for the query performed.}

\textcolor{blue}{This metadata, filled by the authors during their
submission of a preprint, contains different information such as authors,
Arxiv categories and abstract.}

\textcolor{blue}{We process these abstracts using TextBlob, a natural
language processing Python library}\footnote{\textcolor{blue}{\href{https://textblob.readthedocs.io/en/dev/}{https://textblob.readthedocs.io/en/dev/}}}\textcolor{blue}{{}
and extract the nouns using the tagged text.}

\textcolor{blue}{Nouns in the abstract of each document are scored
with TF-IDF, the Term Frequency - Invert Document Frequency, defined
as: 
\[
\text{TF-IDF}\left(x,d\right)=\text{TF\ensuremath{\left(x,d\right)}\ensuremath{\ensuremath{\times\text{IDF\ensuremath{\left(x,d\right)}}}}}
\]
}

\textcolor{blue}{with $\text{TF\ensuremath{\left(x,d\right)}}$ the
relative frequency of $x$ in $d$ and $\text{IDF}\left(x,d\right)$
the invert document frequency.}

\textcolor{blue}{Writing $n_{d}$ the total number of terms in document
$d$ and $n_{x}$ the number of occurences of $x$: 
\[
\text{TF\ensuremath{\left(x,d\right)}}=\dfrac{n_{x}}{n_{d}}
\]
}

\textcolor{blue}{and writing $N$ the total number of documents and
$n_{x\in d}$ the number of documents having an occurence of $x$,
we have}

\textcolor{blue}{
\[
\text{IDF}\left(x,d\right)=\log_{10}\left(\dfrac{N}{n_{x\in d}}\right)
\]
}

\textcolor{blue}{Scoring each noun in each abstract of the retrieved
documents generates a hb-graphs $\mathcal{H}_{Q}$ of universe the
nouns contained in the abstracts. Each hb-edge contains a set of nouns
extracted from a given abstract with a multiplicity function that
represents the TF-IDF score of each noun.}

\begin{figure}
\begin{center}\includegraphics[scale=0.35]{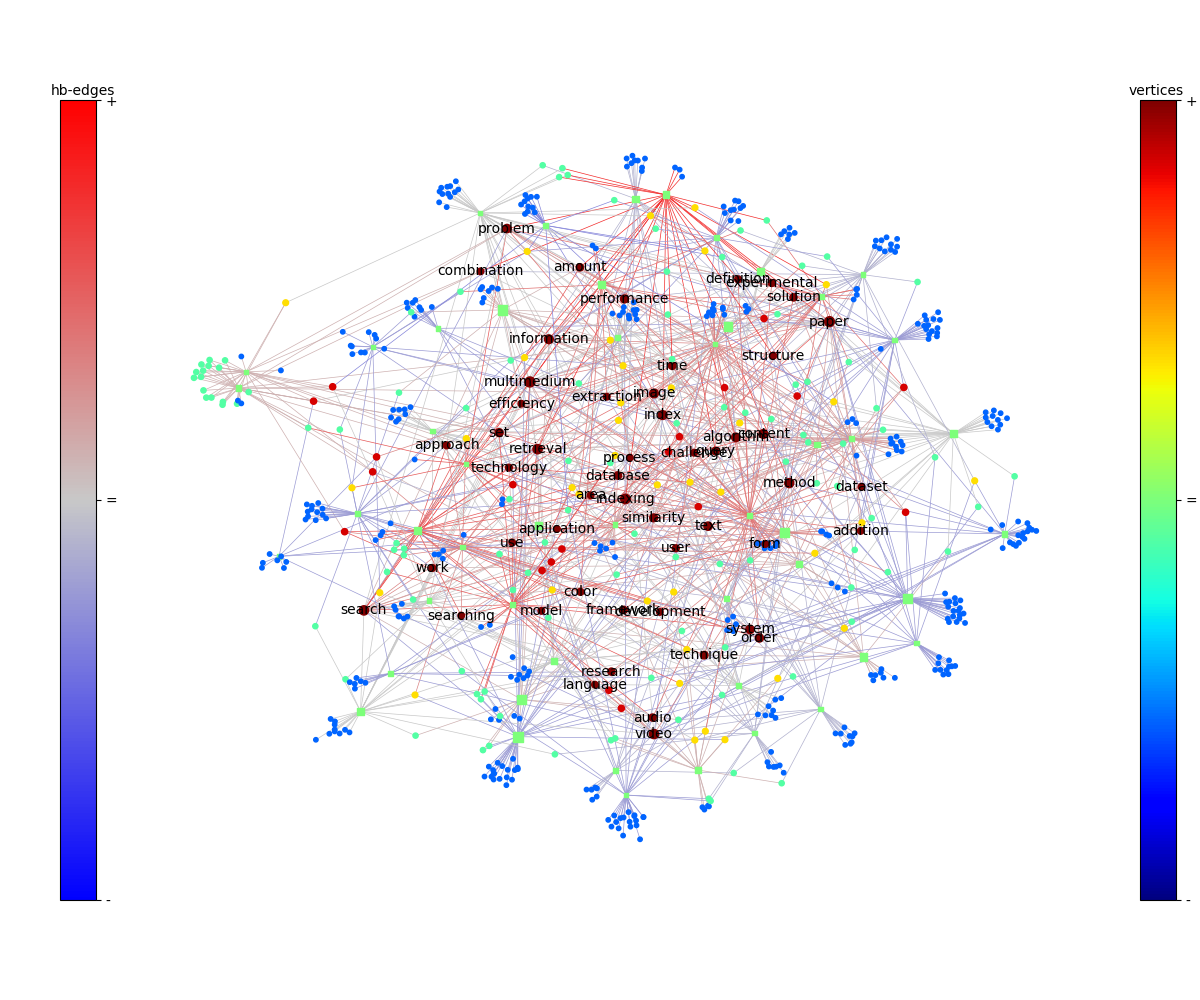}\end{center}

\caption{Querying Arxiv. The search performed is ``content-based multimedia
indexing'' for which 50 most relevant articles have been retrieved
with 50 iterations.}

\label{Fig:Arxiv_cbmi-50}
\end{figure}

\begin{figure}
\begin{center}\includegraphics[scale=0.35]{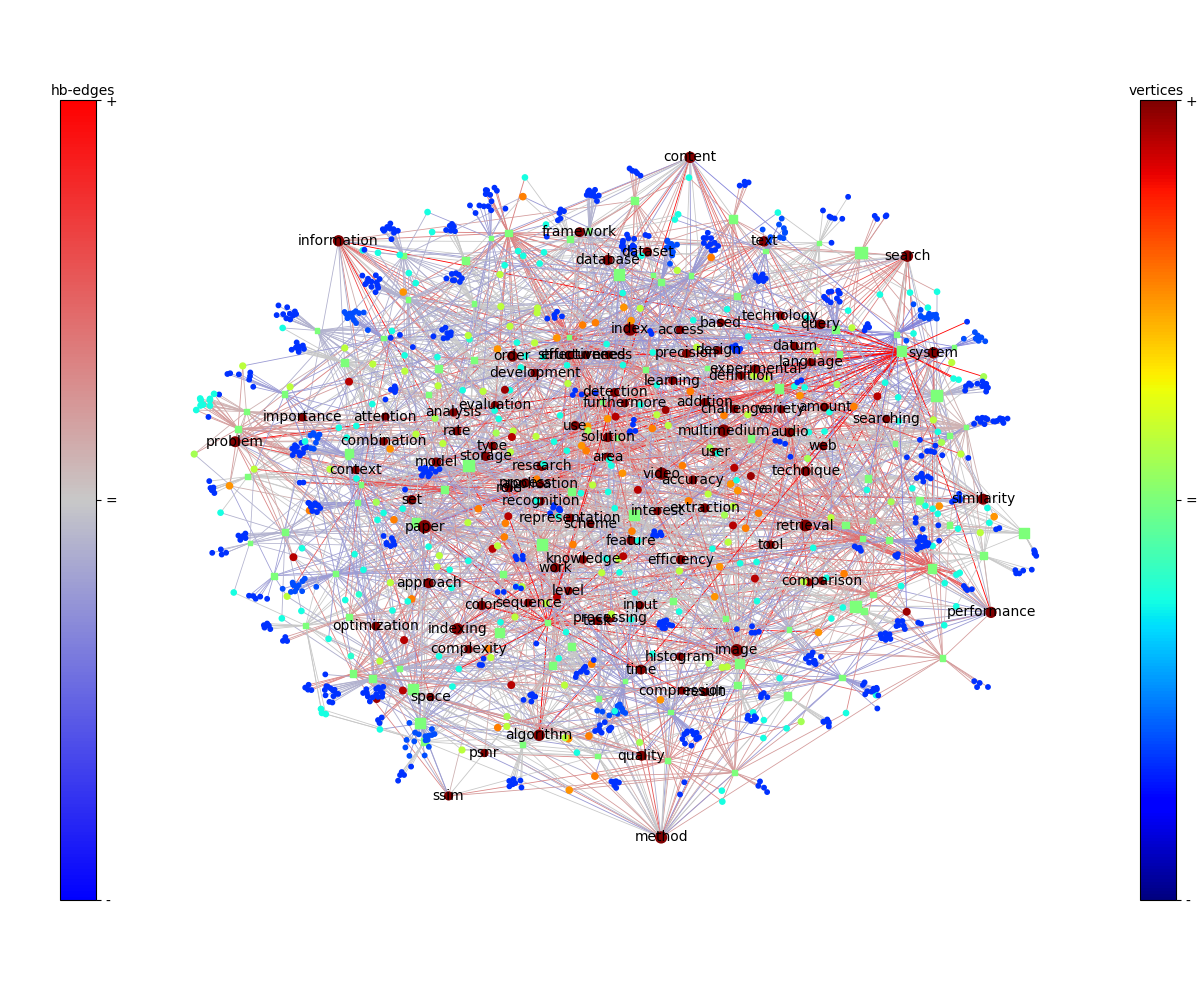}\end{center}

\caption{Querying Arxiv. The search performed is ``content-based multimedia
indexing'' for which 100 most relevant articles have been retrieved}

\label{Fig:Arxiv_cbmi-100}
\end{figure}

\textcolor{blue}{The exchange-based diffusion process is applied to
the hb-graph $\mathcal{H}_{Q}$. We show two typical examples on the
same query the first 50 results in Figure \ref{Fig:Arxiv_cbmi-50}
and the first 100 results in Figure \ref{Fig:Arxiv_cbmi-100}. The
number of iterations needed to have convergence is less than 10 in
these two cases; with 500 results, around 10 iterations are needed
for all hb-edges but one where 30 iterations are needed.}

\textcolor{blue}{As the hb-edges correspond to documents in Arxiv
database we compared the central documents obtained in the results
of the queries: we observe that the ranking obtained based on the
$\epsilon_{49+\frac{1}{2}}$ differs significantly from the ranking
by pertinence given by Arxiv API. In the exchange-based diffusion,
the ranking sorts documents depending on their word weights and their
centrality as we have seen in the experimental part on random hb-graphs.}

\textcolor{blue}{Moreover, we have observed that when the number of
results retrieved increases the top 5, top 10 documents sometimes
change drastically depending on the retrieval of new documents that
are more central in the words they contain. If the gap seems not big
with a few documents retrieved, this gaps increase as the number of
documents increases. The increasing number of results reveal the full
theorical hb-graph obtained from the whole dataset performing the
querying, and hence, reveals central subjects in this dataset. Hence
the diffusion process can allow to highlight importance of documents
by considering central subjects in the processing of the results of
the query.}

\section{Future work and Conclusion}

\label{sec:FW}

The results obtained by using hb-graph highlight the possibility of
using hb-edges for analyzing networks; they confirm that vertices
are highlighted due to their connectivity. The highlighting of the
hb-edges has been achieved by using the intermediate step of our diffusion
process. Different applications can be thought in particular in the
search of tagged multimedia documents for refining the results and
scoring of documents retrieved. Using tagged documents ranking by
this means could help in creating summary for visualisation. Our approach
is seen as a strong basis to refine the approach of \cite{xu2016multi}.
This approach can also be viewed as a mean to make query expansion
and disambiguation by using additional high scored words in the network
and a way of making some recommendation based on the scoring of a
document based on its main words.

\section*{Acknowledgments}

This work is part of the PhD of Xavier OUVRARD, done at UniGe, supervised
by Stéphane MARCHAND-MAILLET and founded by a doctoral position at
CERN, in Collaboration Spotting team, supervised by Jean-Marie LE
GOFF.

\bibliographystyle{ieeetr}
\bibliography{/home/xo/cernbox/these/000-thesis_corpus/biblio/references}

\end{document}